\documentclass{article}
\usepackage{arxiv}
\usepackage[utf8]{inputenc} % allow utf-8 input
\usepackage[T1]{fontenc}    % use 8-bit T1 fonts
\usepackage{url}  
\usepackage{hyperref}       % hyperlinks
\hypersetup{
    colorlinks=true,       % false: boxed links; true:
    linkcolor=blue,          % color of internal links
    citecolor=blue,        % color of links to 
    urlcolor=blue           % color of external links
}
\usepackage{booktabs}       % professional-quality tables math symbols
\usepackage{nicefrac}       % compact symbols for 1/2, etc.
\usepackage{microtype}      % microtypography
\usepackage{lipsum}		% Can be removed after putting your text content
\usepackage{graphicx}
\usepackage{doi}
\usepackage{amsmath}
\usepackage{subfig}
\usepackage[affil-it]{authblk}
\usepackage[english]{babel}
\usepackage{blindtext}
\usepackage{amssymb}
\usepackage{cleveref}
\usepackage{booktabs}
\usepackage{framed,multirow}
\usepackage{hyperref}
\usepackage{url}            % simple URL typesetting
\usepackage{booktabs}% professional-quality tables
\usepackage[mathlines]{lineno}
\usepackage{amsfonts}       % blackboard math symbols
\usepackage{latexsym}
\usepackage{amsmath}
\usepackage{amssymb}
\usepackage{amsfonts}

\usepackage{cleveref}
\usepackage{nicefrac}       % compact symbols for 1/2, etc.
\usepackage{microtype}      % microtypography
\usepackage{lipsum}
\usepackage{graphicx}
\usepackage{adjustbox}
\usepackage{graphicx}
\usepackage[table,xcdraw]{xcolor}
\definecolor{newcolor}{rgb}{.8,.349,.1}
\usepackage[linesnumbered,ruled,vlined]{algorithm2e}

\usepackage{multirow}
\usepackage{diagbox}
\usepackage{subfig}
\usepackage{tikz}
\usetikzlibrary{external}

\usepackage{pgfplots}
\pgfplotsset{compat=1.7}
\usetikzlibrary{positioning}
\usepackage{etoolbox} 
\usepackage[numbers]{natbib}

\newcommand*\linenomathpatch[1]{%
  \cspreto{#1}{\linenomath}%
  \cspreto{#1*}{\linenomath}%
  \csappto{end#1}{\endlinenomath}%
  \csappto{end#1*}{\endlinenomath}%
}
\newcommand*\linenomathpatchAMS[1]{%
  \cspreto{#1}{\linenomathAMS}%
  \cspreto{#1*}{\linenomathAMS}%
  \csappto{end#1}{\endlinenomath}%
  \csappto{end#1*}{\endlinenomath}%
}
\expandafter\ifx\linenomath\linenomathWithnumbers
  \let\linenomathAMS\linenomathWithnumbers
  \patchcmd\linenomathAMS{\advance\postdisplaypenalty\linenopenalty}{}{}{}
\else
  \let\linenomathAMS\linenomathNonumbers
\fi

\linenomathpatch{equation}
\linenomathpatchAMS{gather}
\linenomathpatchAMS{multline}
\linenomathpatchAMS{align}
\linenomathpatchAMS{alignat}
\linenomathpatchAMS{flalign}

\SetCommentSty{mycommfont}

\crefformat{section}{\S#2#1#3}
\crefformat{subsection}{\S#2#1#3}
\crefformat{subsubsection}{\S#2#1#3}
\crefrangeformat{section}{\S\S#3#1#4 to~#5#2#6}
\crefmultiformat{section}{\S\S#2#1#3}{ and~#2#1#3}{, #2#1#3}{ and~#2#1#3}

\date{}
\title{Physics and Equality Constrained Artificial Neural Networks: Application to Forward and Inverse Problems with Multi-fidelity Data Fusion}

\author{\href{https://orcid.org/0000-0002-1095-0881}{\includegraphics[scale=0.08]{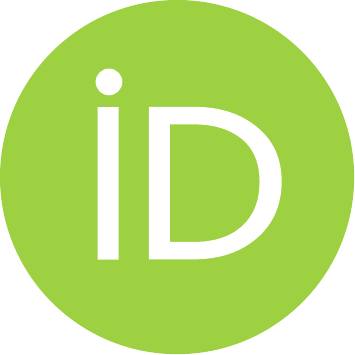}\hspace{1mm}Shamsulhaq Basir}}

\author{\href{https://orcid.org/0000-0003-1967-7583}{\includegraphics[scale=0.08]{orcid.pdf}\hspace{1mm}Inanc Senocak\thanks{corresponding author:~senocak@pitt.edu (Inanc Senocak)}}}

\affil{Department of Mechanical Engineering and Materials Science, University of Pittsburgh, \\ 3700 O'Hara St., Pittsburgh, PA 15261, USA}
\renewcommand{\shorttitle}{Physics and Equality Constrained Artificial Neural Networks}

\begin{document}
\maketitle
\begin{abstract}
Physics-informed neural networks (PINNs) have been proposed to learn the solution of partial differential equations (PDE). In PINNs, the residual form of the PDE of interest and its boundary conditions are lumped into a composite objective function as soft penalties. Here, we show that this specific way of formulating the objective function is the source of severe limitations in the PINN approach when applied to different kinds of PDEs. To address these limitations, we propose a versatile framework based on a constrained optimization problem formulation, where we use the augmented Lagrangian method (ALM) to constrain the solution of a PDE with its boundary conditions and any high-fidelity data that may be available. Our approach is adept at forward and inverse problems with multi-fidelity data fusion. We demonstrate the efficacy and versatility of our physics- and equality-constrained deep-learning framework by applying it to several forward and inverse problems involving multi-dimensional PDEs.
Our framework achieves orders of magnitude  improvements in accuracy levels in comparison with state-of-the-art physics-informed neural networks.
\end{abstract}


\section{Introduction}
 Deep learning has been highly impactful in a plethora of fields such as pattern recognition \cite{krizhevsky2012imagenet,he2016deep}, speech recognition \cite{hinton2012deep},  natural language processing \cite{sutskever2014sequence,bahdanau2016neural,weston2014tagspace} and in the solution of partial differential equations (PDE) for forward and inverse problems. The success of these models owes to the rapid development of available information, the advancement of computing power, and the advent of efficient learning algorithms for training neural networks \cite{schmidhuber2015deep}. With the emergence of universal approximation theorem \cite{hornik1989multilayer,leshno1993multilayer}, new studies have focused on using neural networks to solve ODEs and PDEs. One of the motivations for using neural networks in solving differential equations is their potential to break the curse of dimensionality \cite{grohs2018proof, darbon2020overcoming, berner2020analysis} and its ability to fuse data in the learned solution. Neural network-based methods with their meshless nature can reduce the tedious effort of mesh generation, which is common with finite- difference,  element, or volume methods. Moreover, in contrast to conventional numerical methods, once the neural network is trained, it can produce results at any point in the domain.

 \citeauthor{dissanayake1994neural} pioneered using neural networks to solve PDEs. They combined the residual form of a given PDE and its boundary conditions as soft constraints for training their neural network model. \citeauthor{van1995neural} presented a similar approach and demonstrated its potential on a magnetohydrodynamics plasma equilibrium problem. This general neural network-based technique was applied with satisfactory results to non-linear Schrodinger equations in \cite{monterola2001solving}, to a non-steady fixed bed non-catalytic solid-gas reactor problems in \cite{parisi2003solving}, and to the one-dimensional Burgers equation in \cite{hayati2007feedforward}. A neural network-based approach to solving PDEs and ODEs on orthogonal box domains was also proposed by \citeauthor{lagaris1998artificial} by constructing trial functions that satisfy boundary conditions by construction. Unlike the approach in \cite{dissanayake1994neural, van1995neural}, the approach in \cite{lagaris1998artificial} is limited to regular geometries as it is not trivial to create trial functions for irregular domains. Also, creating trial functions imposes inductive bias toward a certain class of functions that might not be optimal. These early works did not receive broader acceptance and appreciation by other researchers likely because of a lack of computing resources and a limited understanding of neural networks at the time of their introduction.
 
  Machine learning frameworks with automatic differentiation capabilities \cite{abadi2016tensorflow,paszke2019pytorch} have revived the use of neural networks to solve ODEs and PDEs. The overall technical approach for using neural networks to solve PDEs and ODEs that was adopted in the aforementioned works, particularly the method described in \cite{dissanayake1994neural, van1995neural}, has found a resurgent interest in recent years \cite{E2018,Han2018,sirignano2018dgm, Zhu2019}. \citeauthor{Raissi2019} dubbed the term physics-informed neural networks (PINNs), which has been growing fast in popularity and applied to several unique forward and inverse problems \cite{raissi2018deep,raissi2019deep,kissas2020machine,yazdani2020systems,de2020physics,liu2020hierarchical, meng2020composite, Ramabathiran2021}. Even though neural networks offer a powerful framework to faithfully integrate data and physical laws in solving forward and ill-posed inverse problems, training these models is not trivial for challenging problems \cite{van2020optimally,mcclenny2020self,wang2021understanding}. Extensive reviews of the current state in physics-informed machine learning are available in literature \cite{Brunton2020,Karniadakis2021}, but we will also elaborate on the challenges faced by the PINN approach in later sections.
  
   Our paper is structured as follows. In \cref{sec:Tech-Back} we present a technical overview of the physics-based neural networks following the original formulation of \cite{dissanayake1994neural,van1995neural}. Subsequently, we describe a recently proposed empirical algorithm for improving the predictive capability of these models as well as its limitations. Next, we describe the augmented Lagrange method, which forms the backbone of our approach. In \cref{sec:Proposed-Method} we propose the physics and equality constrained artificial neural networks (PECANN) framework and provide a training algorithm for it. In \cref{sec:Forward-Problems} we conduct a comparative analysis of our method on several benchmark problems. In \cref{sec:Inverse-Problems} we demonstrate the performance of the PECANN approach on three different inverse problems with multi-fidelity data fusion. Finally, in \cref{sec:Conclusion} we summarize our results and provide several directions for future research. All the codes and data accompanying this paper are publicly available at \url{https://github.com/HiPerSimLab/PECANN}
\section{Technical Background}
\label{sec:Tech-Back}
Consider a scalar function \(u(\boldsymbol{x},t): \mathbb{R}^{d+1} \rightarrow \mathbb{R}\) on the domain \(\Omega \subset \mathbb{R}^d\) with its boundary \(\partial \Omega\) satisfying the following partial differential equation 
\begin{align}
    \mathcal{F}(\boldsymbol{x},t;\frac{\partial u}{\partial t}, \frac{\partial^2 u}{\partial t^2},\cdots,\frac{\partial u}{\partial \boldsymbol{x}}, \frac{\partial^2 u}{\partial \boldsymbol{x}^2},\cdots,\boldsymbol{\nu}) &= 0,\quad\forall (\boldsymbol{x},t) \in \mathcal{U},\label{eq:PDE}\\
    \mathcal{B}(\boldsymbol{x},t,g;u, \frac{\partial u}{\partial \boldsymbol{x}},\cdots) &= 0, \quad\forall (\boldsymbol{x},t) \in \partial \mathcal{U},\label{eq:BC}\\
    \mathcal{I}(\boldsymbol{x},t,h;u,\frac{\partial u}{\partial t},\cdots) &= 0, \quad \forall (\boldsymbol{x},t) \in \Gamma, \label{eq:IC}
\end{align}
where $\mathcal{F}$ is the residual form of the PDE containing differential operators, $\boldsymbol{\nu}$ is a vector PDE parameters, $\mathcal{B}$ is the residual form of the boundary condition containing a source function $g(\boldsymbol{x},t)$ and $\mathcal{I}$ is the residual form of the initial condition containing a source function $h(\boldsymbol{x},t)$. $\mathcal{U} = \{(\boldsymbol{x},t) ~|~\boldsymbol{x} \in \Omega, t = [0,T] \}$, $\partial \mathcal{U} = \{(\boldsymbol{x},t) ~|~\boldsymbol{x} \in \partial \Omega, t = [0,T]\}$ and $\Gamma = \{(\boldsymbol{x},t) ~|~\boldsymbol{x} \in \partial \Omega, t = 0\}$.
\subsection{Physics-informed Neural Networks}
Here, we present the common elements of the physics-informed learning framework that was presented in the works of \citeauthor{dissanayake1994neural} and \citeauthor{van1995neural}, and, in the work of \citeauthor{Raissi2019} as part of contemporary developments in physics based deep learning methods. Suppose we seek a solution $u_{\theta}(\boldsymbol{x})$ represented by a neural network parameterized by $\theta$ for Eq.~\eqref{eq:PDE} with its boundary condition Eq.~\eqref{eq:BC} and its initial condition Eq.~\eqref{eq:IC}. We can write the following loss functional $\mathcal{L}(\theta)$ to train a physics-informed neural network. 
\begin{align}
    \mathcal{L}(\theta) &=\lambda_{\mathcal{F}}\mathcal{L}_{\mathcal{F}}(\theta) + \lambda_{\mathcal{B}}\mathcal{L}_{\mathcal{B}}(\theta) + \lambda_{\mathcal{I}}\mathcal{L}_{\mathcal{I}}(\theta),\label{eq:PINN_loss}\\
    \mathcal{L}_{\mathcal{F}}(\theta) &= \frac{1}{N_{\mathcal{F}}}\sum_{i=1}^{N_{\mathcal{F}}}\|\mathcal{F}(\boldsymbol{x}^{(i)},t^{(i)})\|_2^2,\label{eq:PDE_Loss}\\
    \mathcal{L}_{\mathcal{B}}(\theta) &= \frac{1}{N_{\mathcal{B}}}\sum_{i=1}^{N_{ \mathcal{B}}}\|\mathcal{B}(\boldsymbol{x}^{(i)}, t^{(i)}, g^{(i)})\|_2^2,\\
    \mathcal{L}_{\mathcal{I}}(\theta) &= \frac{1}{N_{\mathcal{I}}}\sum_{i=1}^{N_{\mathcal{I}}}\|\mathcal{I}(\boldsymbol{x}^{(i)},t^{(i)},h^{(i)})\|_2^2,
\end{align}
where $\{\boldsymbol{x}^{(i)},t^{(i)}\}_{i=1}^{N_{\mathcal{F}}}$ is the set of residual points in $\mathcal{U}$ for approximating the physics loss $\mathcal{L}_\mathcal{F}(\theta)$, $(\{\boldsymbol{x}^{(i)},t^{(i)}),g^{(i)}\}_{i=1}^{N_{\mathcal{B}}}$ is the set boundary points on $\mathcal{\partial \mathcal{U}}$ for approximating the boundary loss $\mathcal{L}_{\mathcal{B}}(\theta)$ and $\{(\boldsymbol{x}^{(i)},t^{(i)}),h^{(i)}\}_{i=1}^{N_{\mathcal{I}}}$ is the set of initial data on $\Gamma$ for approximating the loss on initial condition $\mathcal{L}_{\mathcal{I}}(\theta)$. $\lambda_{\mathcal{F}}$, $\lambda_{\mathcal{B}}$ and $\lambda_{\mathcal{I}}$ are hyperparameters to balance the interplay between the loss terms and $\mathcal{L}(\theta)$ is the sum of all the objective functions used for training a neural network model. It is worth noting that in conventional PINNs $ \lambda_{\mathcal{F}} = \lambda_{\mathcal{B}} = \lambda_{\mathcal{I}} =1$.

Since training PINNs minimizes a weighted sum of several objective functions as in Eq.~\eqref{eq:PINN_loss}, the prediction of the network highly depends on the choice of these weights. Manual setting of these weights by trial and error tuning is extremely tedious and time-demanding. Based on our own experience, we find that manual tuning of these weights is not ideal, because it creates a ripple effect as we then need to tune other hyperparameters, such as the number of collocations points, the learning rate, and the architecture. Also, the optimal choice of these weights for a problem under a certain training setting might not transfer across different problems and may not even produce acceptable results if the training setting is changed. Proper choice of these free parameters is still an active area of research \cite{wang2021understanding,mcclenny2020self,van2020optimally}. Next, we discuss an empirical algorithm proposed by \citeauthor{wang2021understanding} for choosing these hyperparameters.
\subsection{Learning Rate Annealing for Physics-Informed Neural Networks}
 
Consider a physics-informed neural network with parameters $\theta$ and a loss function as follows 
\begin{equation}
    \mathcal{L}(\theta) = \lambda_{\mathcal{F}}\mathcal{L}_{\mathcal{F}}(\theta) + \sum_{i=1}^{M} \lambda_i \mathcal{L}_i(\theta),
    \label{eq:general_pinn_loss}
\end{equation}
where $\mathcal{L}_{\mathcal{F}}(\theta)$ is the PDE residual loss as in Eq.~\eqref{eq:PDE_Loss}, $\mathcal{L}_i(\theta)$ correspond to data-fit terms (e.g., measurements, initial or boundary conditions), $\lambda_{\mathcal{F}}$ and $\lambda_i, i=1,\cdots,M$ are free parameters used to balance the interplay between different loss terms. The necessary optimality condition for Eq.~\eqref{eq:general_pinn_loss} is 
\begin{equation}
    \nabla_{\theta}\mathcal{L}(\theta) = \lambda_{\mathcal{F}}\nabla_{\theta}\mathcal{L}_{\mathcal{F}}(\theta) + \sum_{i=1}^{M} \lambda_i \nabla_{\theta}\mathcal{L}_i(\theta) = 0,
    \label{eq:optimality_condition}
\end{equation}
where $\lambda$s are learned such that the optimality condition is satisfied.
\citeauthor{wang2021understanding} recently proposed an empirical algorithm for setting these weights based on matching the magnitude of the back-propagated gradients as follows
\begin{subequations}
\begin{align}
    \lambda_{\mathcal{F}} &= 1,\\
    \hat{\lambda}_{i} &= \frac{\max_{\theta_n} \{|\nabla_{\theta}\mathcal{L}_{\mathcal{F}}(\theta_n)|\}}{\overline{|\nabla_{\theta_n}\lambda_i \mathcal{L}_i(\theta_n)}|},~ i = 1,\cdots,M,\label{eq:learningRateAnnealing}\\
    \lambda_i &= (1 -\alpha) \lambda_i + \alpha \hat{\lambda_i}, 
\end{align}
\end{subequations}
where $\theta_n$ denotes the values of the network parameters at $n$th iteration, $| \cdot |$ denotes the elementwise absolute value, and the overbar signifies the algebraic mean of the gradient vector. Although this method improves on the original PINN approach $(\lambda_{\mathcal{F}} = \lambda_{i} = 1, i=1,\cdots, M)$, there are fundamental issues with this approach. First, approximating $\hat{\lambda}_i$ in Eq.~\eqref{eq:learningRateAnnealing} does not necessarily meet the optimality condition as in Eq.~\eqref{eq:optimality_condition}. Therefore, the optimizer may settle to a point in the space of parameters that may not be an actual local minimum for the objective function as in Eq.~\eqref{eq:general_pinn_loss}. Second, the values of the network parameters can oscillate back and forth around a minima, which requires slowing down the parameter update by decreasing the learning rate \cite{zeiler2012adadelta}. However, $\hat{\lambda_i}$ grows unbounded when the denominator in Eq.~\eqref{eq:learningRateAnnealing} approaches zero which makes the effective learning rate extremely high and causes the optimizer to diverge. Also, in the case of noisy measurement data, this algorithm tries to fit the noise in the objective function as it is agnostic to the quality of data, and because of the noise in the objective function, its approximated free parameter will oscillate which could hinder convergence. Finally, the method is computationally expensive as it requires $M+1$ number of backward passes through the computational graph to evaluate the gradients of the network parameter with respect to each term in the objective function.
\subsection{Augmented Lagrangian Method for Constrained Optimization}

Consider the following nonlinear, equality-constrained optimization problem with $n$ decision variables, and $m$ equality constraints
\begin{equation}
\begin{split}
&\underset{\boldsymbol{\theta} \in \mathbb{R}^n}{\text{min}}  \hspace{1em} \mathcal{J}(\theta),\\
&\text{subject to} \hspace{1em}\mathcal{C}_{i}(\theta) = 0.~\forall i = 1,\cdots,m
\end{split}
\label{eq:generic_optim_problem}
\end{equation}
where $\mathcal{J}$ is a nonlinear function of $\mathbb{R}^n$ in $\mathbb{R}$, 
$\mathcal{C}_i$ is a nonlinear function of $\mathbb{R}^n$ in $\mathbb{R}$  and $\theta$ is a given subset of $\mathbb{R}^n$, n-dimensional Euclidean space. Augmented Lagrangian method (ALM) \cite{powell1969method,hestenes1969multiplier} which is also the method of choice in the present work can be used to convert the constrained optimization problem of Eq.~\eqref{eq:generic_optim_problem} into an unconstrained optimization problem as follows
\begin{align}
\underset{\theta \in \Theta}{\text{min}}  \hspace{1em}
\mathcal{L}(\theta ;\lambda,\mu) &= \mathcal{J}(\theta) +
\sum_{i=1}^{m} \lambda_i \mathcal{C}_i(\theta) + 
\frac{\mu}{2} \sum_{i=1}^{m} |\mathcal{C}_{i}(\theta)|^2,
\label{eq:augmented_lagrange_method}
\end{align}
where $\lambda \in \mathbb{R}^m$ is a vector of Lagrange multipliers and $\mu$ is a positive penalty parameter, and the semicolon denotes that $\lambda$ and $\mu$ are fixed. We update the vector of Lagrange multipliers based on the
current estimate of the Lagrange multipliers and constraint values using the following rule
\begin{equation}
    \lambda_{i} \leftarrow \lambda_{i} + \mu \mathcal{C}_{i}(\theta).
\end{equation}
In ALM, the objective function is minimized possibly by violating the constraints. Subsequently, the feasibility is restored progressively as the iterations proceed \cite{bierlaire2015optimization}. If $\lambda$ vanish, the penalty method is recovered, whereas when $\mu$ vanishes we get the method of Lagrange multipliers. As discussed in \citeauthor{martins2021engineering}, ALM avoids the ill-conditioning issue of the penalty method while having a better convergence rate than the Lagrange multiplier method \cite{boyd2011distributed}. Therefore, we could say that ALM combines the merit of both methods. Convergence in ALM may occur with finite $\mu$, and optimization problem does not even have to possess a locally convex structure \cite{bertsekas1976multiplier,nocedal2006numerical,boyd2011distributed,bierlaire2015optimization,martins2021engineering}. These aspects of the ALM make it a suitable choice for neural networks as their objective functions are typically non-convex with respect to the parameters of the network. 
ALM has been used in scientific machine learning in the context of PDE-constrained optimization \cite{dener2020training,lu2021physics}. In \citeauthor{dener2020training}, authors train a physics-constrained encoder-decoder neural network using ALM in a supervised learning fashion. In \citeauthor{lu2021physics}, the authors use ALM to train a PDE-constrained neural network model that satisfies the boundary conditions by construction, following an approach similar to the one proposed in \cite{lagaris1998artificial}. 
\section{Proposed Method: Physics \& Equality Constrained Artificial Neural Networks}
\label{sec:Proposed-Method}
Here, we propose a novel approach in using neural networks for the solution of forward problems and inverse problems with multi-fidelity data. This framework is noise-aware, physics-informed and equality constrained. We start by presenting a constrained optimization problem aimed at minimizing the sum of physics loss and noisy data (low-fidelity) loss such that any high fidelity data (boundary conditions, known equality constraints) are strictly satisfied. Considering Eq.\eqref{eq:PDE} with its boundary condition \eqref{eq:BC} and initial condition \eqref{eq:IC}, we write the following constrained optimization problem: 
\begin{align}
    \min_{\theta} \mathcal{J}_{\mathcal{F}}(\theta) + \mathcal{J}_{\mathcal{M}}(\theta),
    \label{eq:objective}
\end{align}
subject to 
\begin{align}
    \phi(\mathcal{B}(\boldsymbol{x}^{(i)},t^{(i)},g^{(i)})) &= 0,~ \forall (\boldsymbol{x}^{(i)},t^{(i)},g^{(i)}) \in \mathcal{\partial U},~ i = 1,\cdots, N_{\mathcal{B}}
    \label{eq:BCconstraints}\\
    \phi(\mathcal{I}(\boldsymbol{x}^{(i)},t^{(i)},h^{(i)})) &= 0,~ \forall (\boldsymbol{x}^{(i)},t^{(i)},h^{(i)}) \in \Gamma,~ i = 1,\cdots, N_{\mathcal{I}},
    \label{eq:ICconstraints}
\end{align}
where $\mathcal{J}_{\mathcal{F}}(\theta)$ is the loss function for the given PDE,  $\phi$ is a distance function and $\mathcal{J}_{\mathcal{M}}(\theta)$ is the objective function for noisy (low-fidelity) measurement data given
\begin{align}
    \tilde{u}(\boldsymbol{x}^{(i)},t^{(i)}) = u_{\theta}(\boldsymbol{x}^{(i)},t^{(i)}) + \epsilon^{(i)},\forall i = 1,\cdots, N_{\mathcal{M}}
    \label{eq:noisy_data}
\end{align}
where $N_{\mathcal{M}}$ is the number of observations, $\tilde{u}(\boldsymbol{x}^{(i)},t^{(i)})$ is the $i$th measurement at $(\boldsymbol{x}^{(i)},t^{(i)})$, $u_{\theta}(\boldsymbol{x}^{(i)},t^{(i)})$ is $i$th prediction from our neural network model at $(\boldsymbol{x}^{(i)},t^{(i)})$ and $\epsilon^{(i)}$ captures the error associated with the $i$th data point. Assuming that the errors are normally distributed with mean zero and a standard deviation of $\sigma$, we can minimizing the \emph{log likelihood} of the predictions $u_{\theta}(\boldsymbol{x},t)$ conditioned on the observed data $\tilde{u}_{\theta}(\boldsymbol{x},t)$ to obtain $\mathcal{J}_{\mathcal{M}}(\theta)$ as follows \cite{martins2021engineering}
\begin{equation}
    \mathcal{J}_{\mathcal{M}}(\theta) =\frac{1}{2\sigma^2} \sum_{i=1}^{N_{\mathcal{M}}}\|u_{\theta}(\boldsymbol{x}^{(i)},t^{(i)}) - \Tilde{u}(\boldsymbol{x}^{(i)},t^{(i)})\|_2^2.
\end{equation}
In this work, we set $\sigma= 1/\sqrt{2} \approx 0.7$ which results in a sum-of-squared errors for the noisy data, however, the user can assign any value to $\sigma$ depending on the quality of the measurement data. It is worth noting, that a smaller value of $\sigma$ which corresponds to less noisy data will put more weight on $\mathcal{J}_{\mathcal{M}}$ and vice versa. Using the augmented Lagrange method, we can write the resulting objective function as follows
\begin{align}
    \mathcal{L}(\theta;\lambda, \mu) &=
    \mathcal{J}_{\mathcal{F}}(\theta) +
    \mathcal{J}_{\mathcal{M}}(\theta) +
    \sum_{i=1}^{N_{\mathcal{B}}} \lambda_{\mathcal{B}}^{(i)} \phi(\mathcal{B}(\boldsymbol{x}^{(i)},t^{(i)},g^{(i)})) + \sum_{i=1}^{N_{\mathcal{I}}} \lambda_{\mathcal{I}}^{(i)} \phi(\mathcal{I}(\boldsymbol{x}^{(i)},t^{(i)},h^{(i)})) + \frac{\mu}{2}\pi(\theta),\label{eq:ProposedObjectiveFunction}\\
    \pi(\theta) &=\sum_{i=1}^{N_{\mathcal{B}}} | \phi(\mathcal{B}(\boldsymbol{x}^{(i)},t^{(i)},g^{(i)}))|^2 + \sum_{i=1}^{N_{\mathcal{I}}} | \phi(\mathcal{I}(\boldsymbol{x}^{(i)},t^{(i)},h^{(i)}))|^2,\\
    \mathcal{J}_{\mathcal{F}}(\theta) &= \sum_{i=1}^{N_{\mathcal{F}}}\|\mathcal{F}(\boldsymbol{x}^{(i)},t^{(i)})\|_2^2,
\end{align}
where $N_{\mathcal{F}}$, $N_{\mathcal{B}}$, $N_{\mathcal{I}}$ are the number of data points in $\mathcal{U}$, $\partial \mathcal{U}$ and $\Gamma$ respectively. We note that any equality constraints can be incorporated as Eq.~\eqref{eq:BCconstraints} and \eqref{eq:ICconstraints} should they arise. $\lambda_{\mathcal{B}} \in \mathbb{R}^{N_{\mathcal{B}}}$ is an $N_{\mathcal{B}}$-dimensional vector of Lagrange multipliers for the constraints on $\partial \mathcal{U}$,
$\lambda_{\mathcal{I}} \in \mathbb{R}^{N_{\mathcal{I}}}$ is an $N_{\mathcal{I}}$-dimensional vector of Lagrange multipliers for the constraints on $\Gamma$,
and $\mu$ is a positive penalty parameter. We update the vector of Lagrange multipliers using the following rule
\begin{align}
    \lambda_{\mathcal{B}}^{(i)} & \leftarrow \lambda_{\mathcal{B}}^{(i)} + \mu 
\phi(\mathcal{B}(\boldsymbol{x}^{(i)},t^{(i)},g^{(i)})), ~\forall (\boldsymbol{x}^{(i)},t^{(i)},g^{(i)}) \in \partial \mathcal{U}, i = 1,\cdots N_{\mathcal{B}},\\
 \lambda_{\mathcal{I}}^{(i)} & \leftarrow \lambda_{\mathcal{I}}^{(i)} + \mu 
\phi(\mathcal{I}(\boldsymbol{x}^{(i)},t^{(i)},h^{(i)})), ~\forall (\boldsymbol{x}^{(i)},t^{(i)},h^{(i)}) \in \Gamma, i = 1,\cdots N_{\mathcal{I}}
\label{eq:lagrange_multiplier_update_rule}
\end{align}

% training algoritm
\IncMargin{1em}
\begin{algorithm}
\SetAlgoLined
% setting keywords 
\SetKw{KwInput}{Input:}
\SetKw{KwOutput}{Output:}
% algorithm 
\KwInput{$\theta^0, \mu_{max},E,S$}\\
$\lambda_{\mathcal{B}}, \lambda_{\mathcal{I}}\leftarrow 0 $ 
\tcc{Initializing the multipliers}
$ \epsilon  \leftarrow 10^{-8}$
\tcc{Assigning the tolerance for constraints violation}
$ \mu_0 \leftarrow 1.0$
\tcc{Initializing the penalty term}
$ \eta \leftarrow 0$\tcc{Placeholder for violation of constraint}
\KwOutput{$\theta^*$}\\
\BlankLine
\For{$epoch \leftarrow 1$ \KwTo $E$}{
    \tcc{Iterate over all training batches}
    \For{$batch \leftarrow 1$ \KwTo $S$}{
        $\theta^* \leftarrow \underset{\theta}{\mathrm{argmin}}~ \mathcal{L}(\theta;\lambda,\mu)$\tcc{Optimizing the network's parameters
        }\uIf{($\sqrt{\pi(\theta)} \hspace{0.20em} \ge \hspace{0.20em} 0.25\eta ) \hspace{0.30em} \& \hspace{0.30em} (\sqrt{\pi(\theta)} \hspace{0.20em}> \hspace{0.20em}\epsilon)$}{
            $\mu \leftarrow \min(2\mu, \mu_{max})$
            \tcc{Updating the penalty parameter}
            $\lambda_{\mathcal{B}} \leftarrow \lambda_{\mathcal{B}} + \phi(\mathcal{B}(\boldsymbol{x},t,g))$\tcc{Updating the Lagrange multiplier for the boundary condition}
            $\lambda_{\mathcal{I}} \leftarrow \lambda_{\mathcal{I}} + \phi(\mathcal{I}(\boldsymbol{x},t,h))$
            \tcc{Updating the Lagrange multipliers for the initial condition}
            }
        $\eta =\sqrt{\pi(\theta)}$\tcc{Recording the current penalty loss}
    }
}
 \caption{Training algorithm for the PECANN framework}
 \label{alg:TrainingAlgorithm}
\end{algorithm}

 In Algorithm \ref{alg:TrainingAlgorithm}, we present a training algorithm using the objective function presented in \eqref{eq:ProposedObjectiveFunction}. 
The input to the algorithm is an initialized set of parameters for the network, a maximum value $\mu_{max}$ for safeguarding the penalty term, the number of epochs $E$, and the training set $S$.
We should note that over-focusing on the constraints might result in a trivial prediction, where the constraints are satisfied, but the solution has not been found. Therefore, we tackle this issue by updating the multipliers when two conditions are met simultaneously: First, the ratio of the penalty loss term from successive iterations has not decreased. Second, the maximum allowable violation on the constraints has not been met. The first condition helps prevent aggressive updating of multipliers that might cause the aforementioned issue. In the second condition, we relax updating the multipliers if a satisfactory precision set by the user $\epsilon$ has been achieved. This, in return, enables the network to freely choose to optimize any loss terms in the objective function to not sacrifice any loss term.

\begin{figure}[!hb]
    \centering
    \subfloat[]{\includegraphics[scale=0.35]{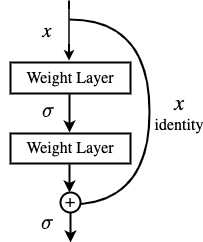}}\quad
    \subfloat[]{\includegraphics[width=0.80\textwidth]{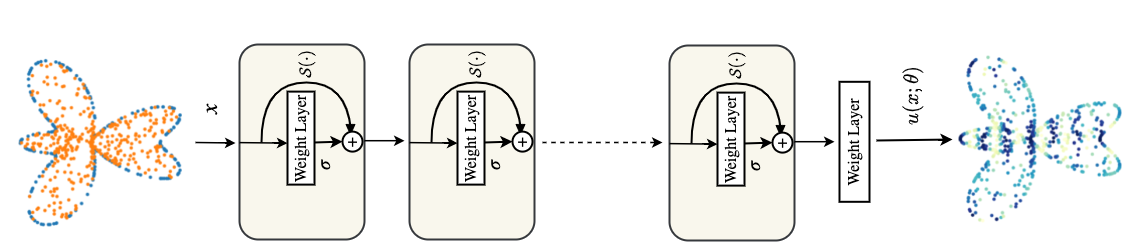}}
    \caption{(a) a schematic representation of the original residual block with a set of parameters $\theta$ and nonlinear activation functions $\sigma$, (b) a schematic representation of our proposed residual neural network architecture with a set of parameters $\theta$ and nonlinear activation functions $\sigma$ with $\mathcal{S}(\cdot)$ skip connections.}
    \label{fig:ResLayerAndNetwork}
\end{figure}

Next, we discuss a ``lean'' residual neural network that we employ for some of our numerical experiments. Conventional feed-forward neural networks are prone to the notorious problem of vanishing-gradients, which makes learning significantly stiff. \citeauthor{he2016deep} proposed residual learning to alleviate this issue by introducing skip connections. Fig.~\ref{fig:ResLayerAndNetwork}(a) shows a schematic 
 representation of a residual block that has two weight layers and a nonlinear activation function $\sigma$ \cite{he2016deep}.  However, to preclude the problem of vanishing gradient, the non-linearity after the summation junction \textcircled{+} and the shortcut connection should be identity as proposed by \citet {he2016identity} as well as in \cite{weinan2017proposal}. We further observe that the weight layer before the junction becomes redundant because the output of the current residual layer will be fed to another residual layer that processes its input through a weight layer. In other words, linearly stacking two weight layers can be collapsed into a single weight layer. Therefore, we eliminate this extra weight layer and obtain a leaner residual layer. A schematic representation of our proposed modified residual layer is shown in Fig.~\ref{fig:ResLayerAndNetwork}(b) with $\mathcal{S}(x)$ shortcut mappings, which are identity mappings except for the input layer to project the input dimension to the correct dimension of the hidden layers.

 \subsection{Performance Metrics}
 We assess the accuracy of our models by providing the $L_\infty$ and the relative $L_{2}$ errors. 
 Given an $n$-dimensional vector of predictions $\boldsymbol{\hat{u}} \in \mathbf{R}^n$ and an $n$-dimensional vector of exact values $\boldsymbol{u} \in \mathbf{R}^n$, we define the relative $L_2$ norm and $L_{\infty}$ norm as follows:
\begin{align}
     \text{Relative}~ L_{2} = \frac{\|\hat{\boldsymbol{u}} - \boldsymbol{u}\|_2}{\|\boldsymbol{u}\|_2}, \quad
     L_{\infty} = \|\hat{\boldsymbol{u}} - \boldsymbol{u}\|_{\infty}
    \label{eq:relativeL2Error}
\end{align}
where $\| \cdot \|_2$ indicates the Euclidean norm.
%------------ Numerical Experiments -----------
\section{Application to Forward Problems}
\label{sec:Forward-Problems}
We apply our framework to learn the solution of several prototypical partial differential equations (PDE) that appear in computational physics. We also compare our results with existing methods to highlight the marked improvements in accuracy levels. 
\subsection{Two-dimensional Poisson's Equation}
\begin{figure}[!ht]
\centering
\includegraphics[width=\textwidth]{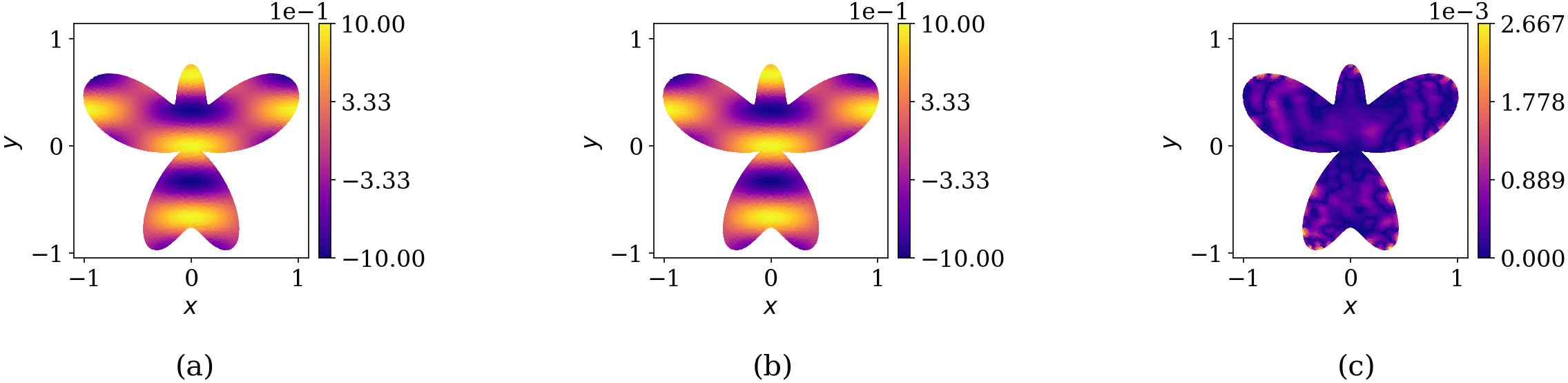}
\caption{Poisson's equation: (a) exact solution, (b) predicted solution by PECANN with quadratic distance function, (c) absolute point-wise error}
    \label{fig:Poisson}
\end{figure}

Elliptic PDEs lack any characteristic path, which makes the solution at every point in the domain influenced by all  other points.  Therefore, learning the solution to elliptic PDEs with neural network based approaches that do not properly constrain the boundary conditions becomes challenging as we will show in this section. Here, we solve a two-dimensional Poisson's equation on a complex domain to not only highlight the applicability of our approach to irregular domains, but also show that our framework properly imposes the boundary conditions and produces physically feasible solutions. We also conduct a study to show the impact of distance functions $\phi$ and the maximum penalty parameter $\mu_{\max}$ that appear in Eq.~\eqref{eq:ProposedObjectiveFunction} on the prediction of our neural network model. Let us consider the following PDE:
\begin{subequations}
\begin{align}
    \nabla^2 u(x,y) &= f(x,y), \qquad (x,y) \in \Omega,\label{eq:HeatEquation}\\
    u(x,y) &= h(x,y) \qquad (x,y)\in \partial \Omega,\label{eq:HeatEquationBC}
\end{align}
\end{subequations}
where $f(x,y)$ and $h(x,y)$ are source functions, $\Omega = \{ (x,y) ~|~  x = 0.55 \rho(\theta) \cos(\theta), ~y= 0.75 \rho(\theta) \sin(\theta)\}$ and $\rho(\theta) = 1 + \cos(\theta)\sin(4 \theta)$ for $0 \le \theta \le 2 \pi$. We manufacture a complex oscillatory solution for Eq.~\eqref{eq:HeatEquation} and its boundary conditions Eq.~\eqref{eq:HeatEquationBC} as follows:
\begin{equation}
        u(x,y) = \cos(\pi x) \cos(3 \pi y), \qquad (x,y)\in \Omega.
        \label{eq:exact_poisson_2d}
\end{equation}
 The corresponding source functions $f(x,y)$ and $g(x,y)$ can be calculated exactly using Eq.~\eqref{eq:exact_poisson_2d}. We use our ``lean'' residual neural network architecture with 3-layer hidden layers and 50 neurons per layer. We generate $N_{\Omega} = 512$ residual points uniformly from the interior part of the domain at each optimization step and $N_{\partial \Omega} = 512$ from the boundaries only once before training. Our optimizer is Adam with its default parameters and an initial learning rate of $10^{-2}$. We train our network for $25000$ epochs. We reduce our learning rate by a factor of $0.95$ after 100 epochs with no improvement using \emph{ReduceLROnPlateau} learning scheduler that is built in PyTorch framework \cite{paszke2019pytorch}. For the present case, the predictions of both models for the entire domain are juxtaposed in Fig.~\ref{fig:Poisson}.
From Fig.~\ref{fig:Poisson} we observe that our neural network model trained with our proposed approach has successfully learned the underlying solution. Since our physics-informed neural network model diverged, we do not portray its prediction for the entire domain. However, we present a summary of our error norms averaged over five independent trials with random Xavier initialization scheme \cite{glorot2010understanding} for both approaches in Table~\ref{tb:Poisson}. The results indicate that our method achieves a relative $L_2 = 5.90 \times 10^{-4}$, which is three orders of magnitude lower than the one obtained from conventional physics-informed neural networks.
\begin{table}[ht!]
\centering
\caption{2D Poisson's equation. Summary of the average and the standard deviation of the relative $L_2$ and $L_{\infty}$ errors over 5 independent trials along with the number of generated collocation points for training a fixed neural network architecture with different methods.}
\label{tb:Poisson}
\vspace{2pt}
\resizebox{0.70\textwidth}{!}{%
\begin{tabular}{@{}lcccccc@{}}
\toprule
\multicolumn{1}{c}{Models} &
  \multicolumn{1}{c}{Relative $L_{2}$} &
  \multicolumn{1}{c}{$L_\infty$} &
    \multicolumn{1}{c}{$N_{\Omega}$} &
  \multicolumn{1}{c}{$N_{\partial \Omega}$} &
  \\ \midrule
PINN  & $1.29 \times 10^{-1} \pm 2.28 \times 10^{-2}$ & $4.67 \times 10^{-1} \pm 8.68 \times 10^{-2}$& $512 \times 25000 $&512\\
PECANN & $\boldsymbol{5.90 \times 10^{-4} \pm 7.69 \times 10^{-5} }$& $\boldsymbol{4.12 \times 10^{-3} \pm 1.47 \times 10^{-3}}$& $512 \times 25000 $ &512
\end{tabular}}
\end{table}

\begin{figure}[!ht]
     \subfloat{\includegraphics[scale=0.45]{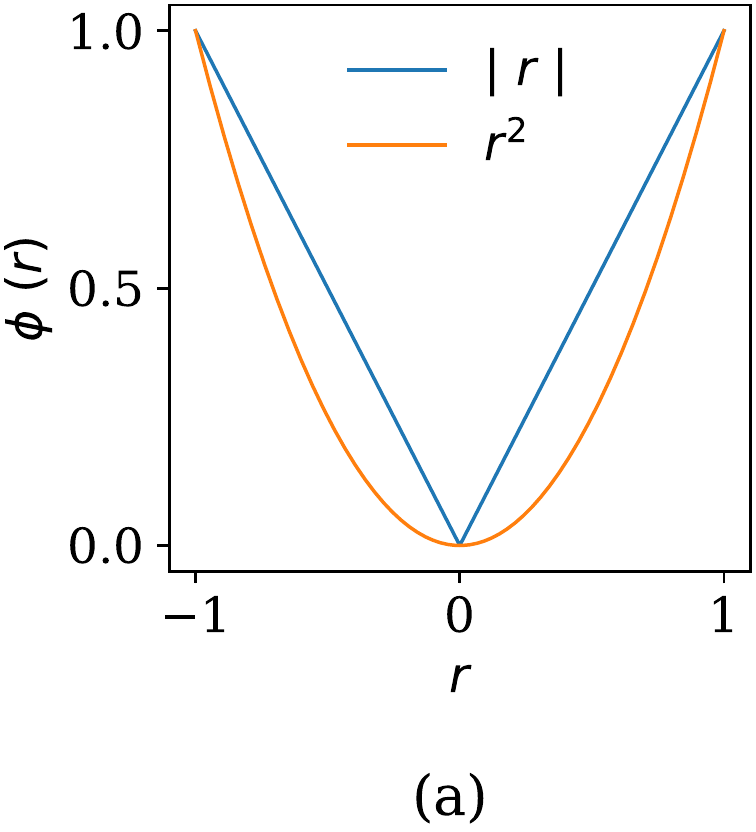}}\hspace{7em}
    \subfloat{\includegraphics[scale=0.45]{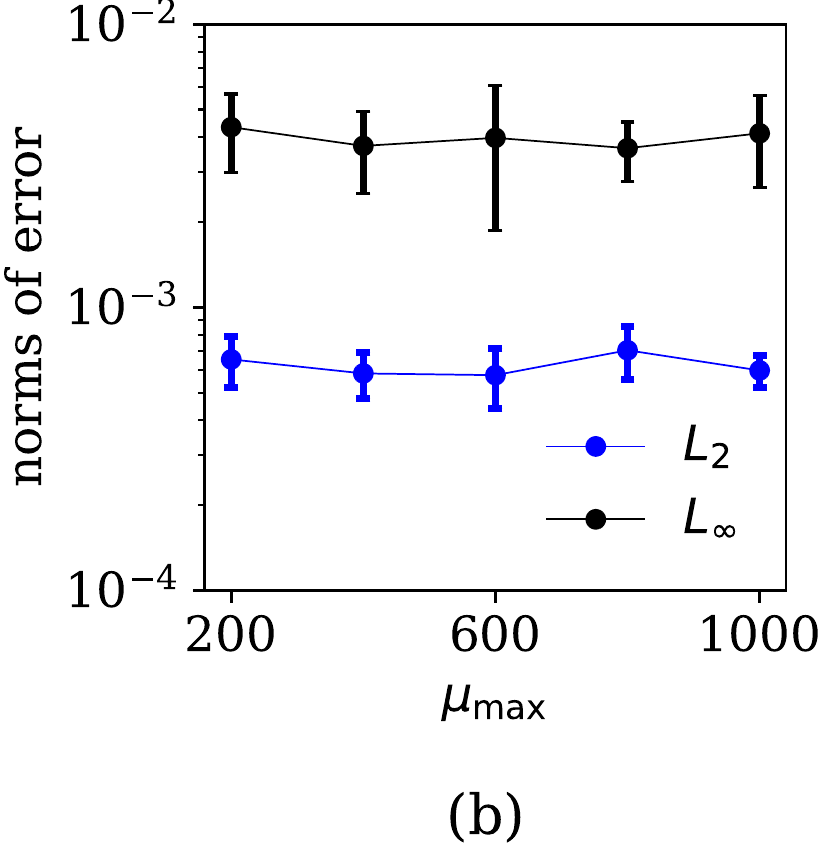}}\hspace{7em}
    \subfloat{\includegraphics[scale=0.45]{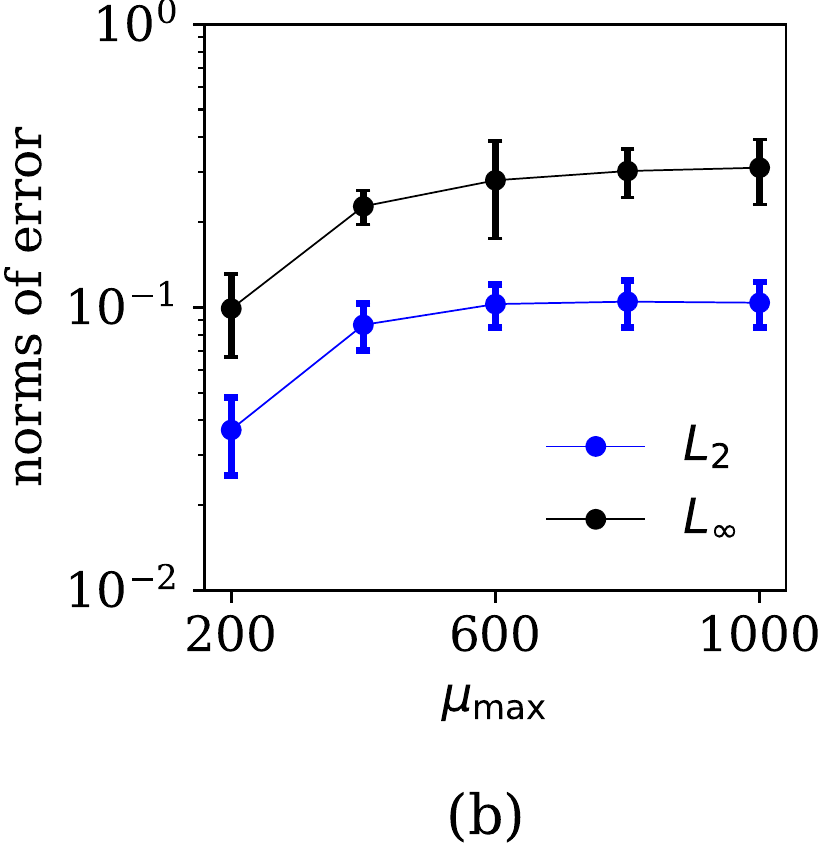}}
     \caption{(a) quadratic and absolute distance functions, (b) relative $L_2$ error bars versus $\mu_{\max}$ for quadratic distance function averaged over 5 independent trials,(c) relative $L_2$ error bars versus $\mu_{\max}$ for absolute distance function averaged over 5 independent trials}
     \label{fig:distance_function}
 \end{figure}

Next, we conduct an ablation study to investigate the impact of the distance function $\phi$ on the prediction of our model. A schematic representation of two different distance functions are presented in Fig.~\ref{fig:distance_function}(a). Our analysis reveals that quadratic distance functions are not only insensitive to the choice of the maximum penalty parameter $\mu_{\max}$ but also significantly outperform the absolute distance function as shown in Fig.~\ref{fig:distance_function}(b)-(c). Therefore, we adopt the quadratic distance function in our proposed method. 
 
To complement our analysis of elliptic PDEs, we present the applications of the PECANNs for a one-dimensional and a three-dimensional Poisson equation in the appendix.

\subsection{Two-dimensional Helmholtz Equation}
Helmholtz equation arises in the study of electromagnetic radiation \cite{li2013three,greengard1998accelerating}, seismology \cite{plessix2007helmholtz}, acoustics \cite{bayliss1985numerical} and many areas of engineering science. In this section, we study the following benchmark problem that was presented in \cite{wang2021understanding}
\begin{subequations}
\begin{align}
        \nabla^2 u(x,y) + k^2 u(x,y) &= q(x,y), ~\forall (x,y) \in \Omega,\\
        u(x,y) &= 0, ~ \forall (x,y) \in \partial \Omega,
        \label{eq:HelmholtzPDE}
\end{align}
\end{subequations}
where $k=1$,  $\Omega = \{ (x,y) ~ | ~ -1 \le x \le 1, -1 \le y \le 1 \}$ and $\partial \Omega$ is its boundary. Following the equation presented above, we manufacture an oscillatory solution that satisfy Eq.~\eqref{eq:HelmholtzPDE} with its boundary conditions as follows:
\begin{equation}
u(x,y) = \sin( \pi x)\sin(4 \pi y), \forall (x,y) \in \Omega.
 \label{eq:HelmholtzExactSolution}
\end{equation}
We use the same fully connected neural network architecture as in \cite{wang2021understanding}, which consists of three hidden layers with 30 neurons per layer and the tangent hyperbolic activation function. We use a Sobol sequence to sample $N_{\Omega} = 512$ residual points from the interior part of the domain and $N_{\partial \Omega} = 256$ from the boundaries only once before training. We note that \cite{wang2021understanding} is generating their data at every epoch, which amounts to $N_{\Omega} = 5.12 \times 10^{6}$ and $N_{\partial \Omega} = 20.48 \times 10^{6}$.  Our optimizer is L-BFGS  \cite{nocedal1980updating} with its default parameters and \emph{strong wolfe} line search function that is built in PyTorch framework \cite{paszke2019pytorch}. We train our network for 5000 epochs with our safeguarding penalty parameter $\mu_{\max} = 10^{4}$.
\begin{figure}[t]
\centering
\includegraphics[width=\textwidth]{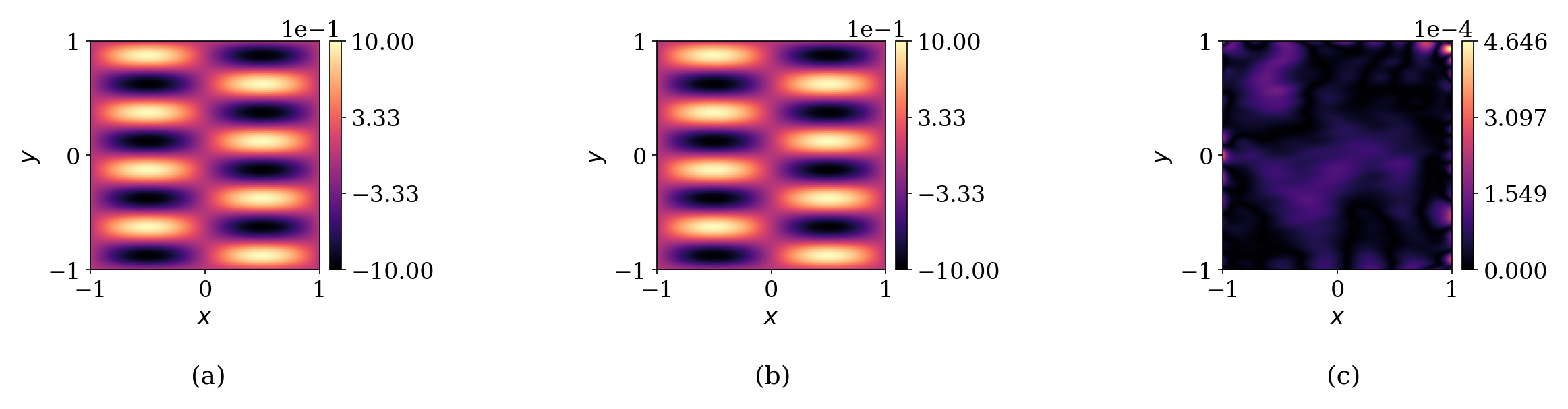}
\caption{Helmholtz equation: (a) exact solution ,(b) predicted solution from PECANN model, (c) absolute point-wise error}
\label{fig:Helmholtz}
\end{figure}
% table
As illustrated in Fig.~ \ref{fig:Helmholtz}(b), our PECANN model produces an accurate prediction to the underlying solution with uniform error distribution across the domain as shown in Fig.~\ref{fig:Helmholtz}(c). We also present a summary of the error norms from our approach and state-of-the-art results presented in \cite{wang2021understanding} averaged over ten independent trials with random Xavier initialization scheme\cite{glorot2010understanding} in Table~\ref{tb:Helmholtz}. We observe that results obtained from our method achieves a relative $L_2 = 4.23 \times 10^{-4}$, which is two orders of magnitude lower than $4.31 \times 10^{-2}$ obtained from the method presented in \citet{wang2021understanding} with only a fraction of their generated data. 
\begin{table}[h]
\centering
\caption{Helmholtz equation: summary of the average and the standard deviations of the relative $L_2$ and $L_{\infty}$ errors over 10 independent trials along with the number of generated collocation points for training a fixed neural network architecture with different methods along}
\label{tb:Helmholtz}
\vspace{2pt}
\resizebox{0.80\textwidth}{!}{%
\begin{tabular}{@{}lccccc@{}}
\toprule
\multicolumn{1}{c}{Models} &
  \multicolumn{1}{c}{Relative $L_{2}$} &
  \multicolumn{1}{c}{$L_\infty$} &
  \multicolumn{1}{c}{$N_{\Omega}$} &
  \multicolumn{1}{c}{$N_{\partial \Omega}$}
  \\ \midrule
Ref. \cite{wang2021understanding}  & $4.31 \times 10^{-2} \pm 1.68 \times 10^{-2}$& - & $128 \times 40000$ & $4 \times 128 \times 40000 $ \\
PECANN  & $\boldsymbol{4.23 \times 10^{-4} \pm 3.09 \times 10^{-4}}$ & $\boldsymbol{1.53 \times 10^{-3} \pm 7.66 \times 10^{-4}}$ & 512 & $4 \times 64$&
\end{tabular}}
\end{table}
\subsection{Klein-Gordon Equation}
We consider a nonlinear time-dependent benchmark problem known as the Klein-Gordon equation, which plays a significant role in many scientific applications such as particle physics, astrophysics, cosmology, and classical mechanics. This problem was considered in the work of \citet{wang2021understanding} as well. Consider the following partial differential equation
\begin{subequations}
\begin{align}
    \frac{\partial^2 u}{\partial t^2 } + \alpha  \frac{\partial^2 u}{\partial x^2 } + \beta u + \gamma u ^ k &= f(x,t), ~\forall (x,t) \in \Omega \times [0,T],\\
    u(x,0)&= g_1(x), ~\forall x \in \Omega,\label{eq:Klein_Gordon_IC1}\\
    \frac{\partial u(x,0)}{\partial x} &= g_2(x),\label{eq:Klein_Gordon_IC2} ~\forall x \in \Omega,\\
    u(x,t) &= h(x,t) ~\forall (x,t) \in \partial \Omega \times [0,T] \label{eq:Klein_Gordon_BC},
\end{align}
\end{subequations}
where $\alpha = -1 $, $\beta = 0$, $\gamma = 1$ and $k= 3$ are known constants. $\Omega =[0,1]\times[0,1]$ with $T=1$. The manufactured solution presented in \cite{wang2021understanding} is as follows 
\begin{equation}
    u(x,t) =  x \cos(5\pi t) + (xt)^{3}.
    \label{eq:Klein_Gordon_exact}
\end{equation}
The corresponding forcing function $f(x,t)$, boundary condition $h(x,t)$ and initial conditions $g_1(x)$ and $g_2(x)$ can be calculated exactly using Eq.~\eqref{eq:Klein_Gordon_exact}. We use the same neural network architecture as in \cite{wang2021understanding} which is a deep fully connected neural network with 5 hidden layers each with 50 neurons that we train for 1500 epochs total. We use Sobol sequence to generate $N_{\Omega} = 512$ residual points from the interior part of the domain, $N_{\partial \Omega} = 512$ points from the boundaries and $N_{I} = 256$ points for each of the initial conditions as in Eq.~\eqref{eq:Klein_Gordon_IC1} and Eq.~\eqref{eq:Klein_Gordon_IC2} only once before training. Our optimizer is LBFGS with its default parameters and \emph{strong wolfe} line search function that is built in PyTorch framework \cite{paszke2019pytorch}. Our safeguarding penalty parameter $\mu_{\max} = 10^{4}$ as in the previous problem.
\begin{figure}[t]
\centering
\includegraphics[width=\textwidth]{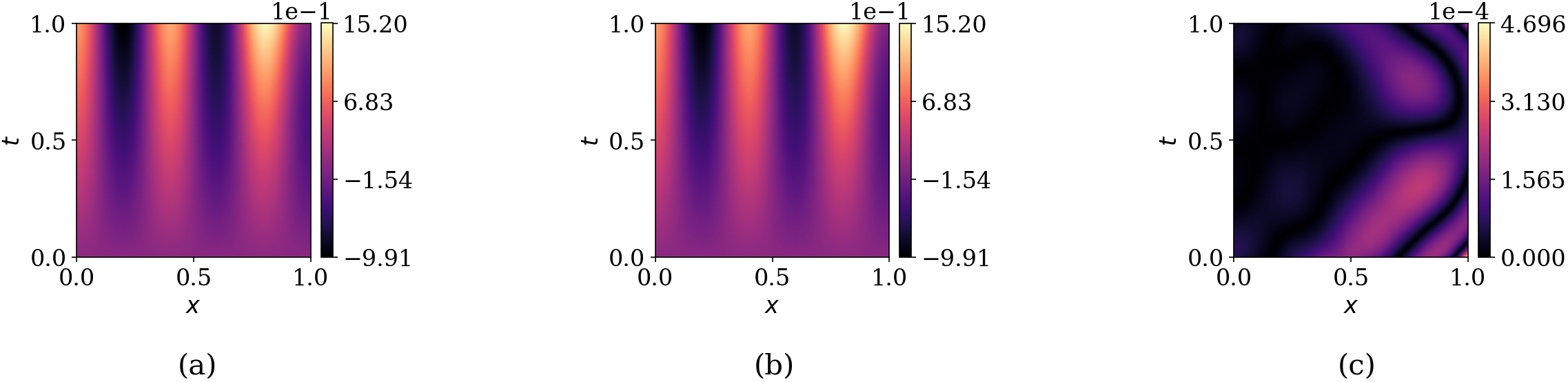}
\caption{Klein Gordon equation: (a) exact solution , (b) predicted solution by PECANN, (c) point-wise absolute error }
\label{fig:Klein_Gordon}
\end{figure}

As illustrated in Fig.~\ref{fig:Klein_Gordon}(b), our PECANN model produces an accurate prediction to the underlying solution with uniform error distribution across the domain as shown in Fig.~\ref{fig:Klein_Gordon}(c). In addition, we present a summary of the error norms averaged over ten independent trials with random Xavier initialization scheme\cite{glorot2010understanding} in Table~\ref{tb:Klein_Gordon}. We observe that the best relative $L_2$ error obtained from our PECANN model is two orders of magnitude lower than the best relative $L_2$ norm error reported in \cite{wang2021understanding} with only a fraction of their generated data. This highlights the predictive power of our method over state-of-the-art physics-informed neural networks for the solution of a non-linear time-dependent Klein-Gordon equation. 
\begin{table}[h]
\centering
\caption{Klein–Gordon equation:summary of the average and the standard deviations of the relative $L_2$ and $L_{\infty}$ errors over 10 independent trials along with the number of generated collocation points for training a fixed neural network architecture with different methods alongs}
\label{tb:Klein_Gordon}
\vspace{2pt}
\resizebox{0.95\textwidth}{!}{%
\begin{tabular}{@{}lcccccc@{}}
\toprule
\multicolumn{1}{c}{Models} &
  \multicolumn{1}{c}{Best Relative $L_{2}$} &
  \multicolumn{1}{c}{Relative $L_{2}$} &
  \multicolumn{1}{c}{$L_\infty$} &
  \multicolumn{1}{c}{$N_{\Omega}$} &
  \multicolumn{1}{c}{$N_{\partial \Omega}$}&
  \multicolumn{1}{c}{$N_{I}$}
  \\ \midrule
Ref. \cite{wang2021understanding}  & $1.062 \times 10^{-2}$& - &-& $128 \times 40000$ & $2 \times 128 \times 40000$ & $ 2 \times  128 \times  40000$ \\
PECANN  & $\boldsymbol{2.158 \times 10^{-4}}$ & $\boldsymbol{6.139 \times 10^{-4} \pm 3.337 \times 10^{-4}}$ & $\boldsymbol{1.043 \times 10^{-3} \pm 5.908 \times 10^{-4}}$& $512$ & $2 \times 256$ & $2 \times 256$
\end{tabular}}
\end{table}
\section{Application to Inverse Problems}
\label{sec:Inverse-Problems}

In this section, we apply our PECANN framework for the solution of inverse problems with multi-fidelity data. By multi-fidelity, we mean that we have both clean (high-fidelity) data and noisy (low-fidelity) data. We tackle three inverse problems involving PDEs. It is worth reiterating that we only impose equality constraints and use noisy data (e.g., noisy boundary conditions, noisy measurement data) as a soft-regularizer $\mathcal{J}_{\mathcal{M}}(\theta)$ in Eq.~\eqref{eq:ProposedObjectiveFunction}. 

\subsection{Learning Hydraulic Conductivity of Nonlinear Unsaturated Flows from Multi-fidelity Data}
Our PECANN framework is also suitable to solve inverse-PDE problems using multi-fidelity data. With multi-fidelity, we mean that the observed data may include both data with low accuracy and data with very high accuracy. As part of our objective function formulation, we can constrain the high-fidelity data in a principled fashion and take advantage of the low-fidelity data to regularize our hypothesis space. To demonstrate our framework, we study one of the difficult multi-fidelity example problems that were tackled in \citeauthor{meng2020composite} with composite neural networks. This particular inverse-PDE problem arises in unsaturated flows as they are central in characterizing contaminant transport \cite{javadi2007flow}, soil-atmosphere interaction \cite{an2017numerical}, soil-plant-water interaction \cite{gadi2018modeling}, ground-subsurface water interaction zone \cite{hayashi2002effects} to name a few. Describing processes involving soil-water interactions at a microscopic level is very complex due to the existence of tortuous, irregular, and interconnected pores \cite{hillel1998environmental}. Therefore, these flows are generally characterized in terms of their macroscopic characteristics. An important quantity that is essential in describing flows through unsaturated soil is hydraulic conductivity, which is a nonlinear parameter that is highly dependent on the geometry of the porous media \cite{hillel1998environmental}. 
Let us consider the following nonlinear differential equation representing an unsaturated one-dimensional (1D) soil column with variable water content:
%%%%
\begin{equation}
    \frac{d}{dx} (K(h) \frac{dh(x)}{dx}) = 0, \qquad x \in \Omega,
    \label{eq:UnsaturatedPDE}
\end{equation}
subject to the following boundary conditions, 
\begin{subequations}
\begin{align}
    h(0)   &= -3,\\
    h(200) &= -10,
\end{align}
\end{subequations}
where $\Omega = \{x ~|~ 0 \le x \le 200 ~\text{cm}\}$ , $h(x)$ is the  pressure head (cm) and $K(h)$ is the hydraulic conductivity (cm $h^{-1}$) which is described as follows:
\begin{equation}
    K(h) = K_s S_e^{1/2}\Big[1 - (1 - S_e^{1/m})^m \Big]^2,
    \label{eq:K(h)}
\end{equation}
where $K_s$ is the saturated hydraulic conductivity (cm $h^{-1}$), and $S_e$ is the effective saturation expressed as follows \cite{van1980closed}: 
\begin{equation}
    S_e = \frac{1}{(1 + |\alpha h|^n)^m}, m = 1 -1/n,
    \label{eq:Se}
\end{equation}
where $\alpha$ is an empirical parameter that is inversely related to
the air-entry pressure value ($\text{cm}^{-1}$ ) and \(m\) is an empirical parameter related to the pore-size distribution that is hard to measure due to the complex geometry of the porous media. We aim to infer the unknown empirical parameters $\alpha$, and $m$ from sparse measurements of pressure head $h$. To generate multi-fidelity synthetic measurements or experimental data, we select the soil type \textit{loam} for which the empirical parameters are as follows: $\alpha = 0.036$ and $m = 0.36$. 

\begin{figure}[!ht]
    \centering
    \includegraphics[width=\textwidth]{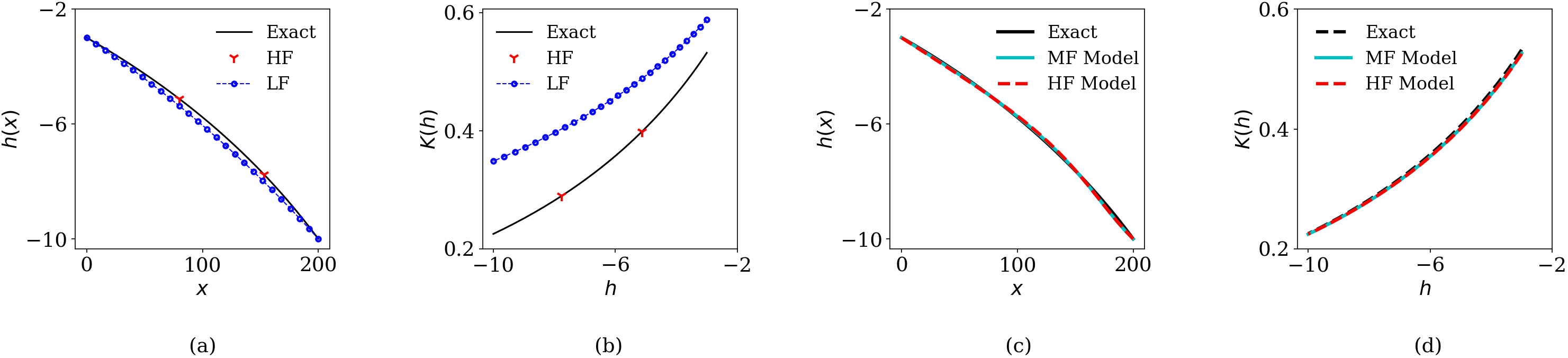}
    \caption{Parameter inference on multi-fidelity data for unsaturated flow through porous media: (a) low fidelity (LF) and high fidelity (HF) pressure head data used for training, (b) hydraulic conductivity corresponding to low-fidelity and high-fidelity training data, (c) pressure head reconstruction by PECANN model trained on high-fidelity and multi-fidelity data separately, (d) reconstructed hydraulic conductivity by PECANN model trained by high-fidelity and multi-fidelity data separately. }
    \label{fig:soil_column}
\end{figure}

We generate high-fidelity pressure data using the exact empirical parameters and low-fidelity data with $\alpha = 0.015$ and $m = 0.31$. Using the built-in \texttt{bvp5c} MATLAB function, we solve the governing PDE as given in Eqs.~\ref{eq:UnsaturatedPDE} through Eq.~\ref{eq:Se} using the selected empirical parameters to generate multi fidelity training data as shown in Fig~\ref{fig:soil_column}(a). In Fig~\ref{fig:soil_column}(b) we also depict the corresponding hydraulic conductivity $k(h)$ values for the pressure head data, which shows that low-fidelity hydraulic conductivity has a significant deviation from the exact hydraulic conductivity distribution. To highlight the robustness, efficiency, and accuracy of our framework on an inverse-PDE with multi-fidelity data fusion, we compare our results with the results reported in \citet{meng2020composite}. For comparison purposes, we also choose a feed-forward neural network with two hidden layers with 20 neurons per layer as in \cite{meng2020composite} for their physics-informed neural network trained on high fidelity alone which failed to discover the parameters of interest. However, \citet{meng2020composite} constructed customized networks for high fidelity data and low fidelity data separately and then aggregated them together by manually crafted correlations. Therefore, they refer to their approach as composite neural networks. Unlike \citet{meng2020composite}, we do not need to make any inductive bias about the data and, therefore, use a single network initialized with Xavier initialization technique \cite{glorot2010understanding} that we separately train on high-fidelity and multi-fidelity data. This shows the robustness and efficiency of our approach that we can train the same network on multi-fidelity data without the need to design customized networks to process data differently. We let a single network discover and extract features from multi-fidelity data with the help of known physics.  We use Adam with its default parameters and $10^{-2}$ initial learning rate. We set the maximum penalty parameter $\mu_{\max} = 10^{4}$ and train our network for 2000 epochs total. 
 As for the collocation points, we use the Sobol sequence and generate 400 residual points from across our domain in each epoch. As considered in \cite{meng2020composite}, we assume the flux at the inlet $q_0$ is known, which allows us to use the integral form of Eq.~\eqref{eq:UnsaturatedPDE} given as follows, 
\begin{equation}
    q(x) = -K(h) \frac{d h(x)}{dx} = q_0, \quad \frac{d q(x)}{dx} = 0.
\end{equation}

Fig.~\ref{fig:soil_column}(a) and Fig.~\ref{fig:soil_column}(b) depict the reconstructed pressure head and the corresponding hydraulic conductivity distributions obtained from our PECAN trained on high-fidelity and multi-fidelity data separately. Compared with the exact solution, it is seen that the inferred results are highly accurate, which shows the robustness and efficiency of our method. Furthermore, in Table~\ref{tb:inverseBenchmarkComparison}, we report the average and standard deviation of inferred $\alpha$ and $m$ from our model along with the results from \citet{meng2020composite}. The results are over 10 independent trials with random initialization using Xavier \cite{glorot2010understanding} scheme. 

From Table~\ref{tb:inverseBenchmarkComparison}, we observe that our results are significantly outperforming the reported results in \cite{meng2020composite}. It is worth noting that we are using just a single neural network architecture that is the low-fidelity model in the composite neural network model proposed in \cite{meng2020composite} and our average CPU training time is only 4 seconds.

\begin{table}[h]
\centering
\caption{ Summary of the inferred parameters from using high-fidelity (HF) only or multi-fidelity (MF) data in the learning process averaged over ten different runs. Note that the training time, averaged over 10 independent trials, for our PECANN model is only 4 seconds on a CPU.}
\label{tb:inverseBenchmarkComparison}
\vspace{2pt}
\resizebox{0.9\textwidth}{!}{%
\begin{tabular}{@{}lccccccc@{}}
\toprule
\multicolumn{1}{c}{Models} &
  \multicolumn{1}{c}{Avg. $\alpha$} &
  \multicolumn{1}{c}{$\sigma(\alpha)$} &
  \multicolumn{1}{c}{Relative Error($\alpha$)}&
  \multicolumn{1}{c}{Avg. $m$} &
  \multicolumn{1}{c}{$\sigma(m)$} &
    \multicolumn{1}{c}{Relative Error($m$)}
  \\ \midrule
Ref. \cite{meng2020composite} with HF data only  & 0.0440  & - & 22.22 \% & 0.377 & - & 4.72 \%\\
PECANN with HF data only  & $\boldsymbol{0.0351}$ & $\boldsymbol{7.18 \times 10^{-4}}$ & $\boldsymbol{2.58 \% }$& $\boldsymbol{0.354}$ &$\boldsymbol{2.78\times 10^{-3}}$ & $\boldsymbol{1.78\%}$
\\ 
Ref. \cite{meng2020composite} with MF data &  0.0337 &  $ 7.91 \times 10^{-4}$ & 6.39\% & 0.349 &$3.70 \times 10^{-3}$& 3.06\% \\
PECANN with MF data  &$\boldsymbol{0.0359}$ & $\boldsymbol{7.51 \times 10^{-4}}$ & $\boldsymbol{0.30\% }$& $\boldsymbol{0.357}$ & $\boldsymbol{2.74\times 10^{-3}}$ & $\boldsymbol{0.86\% }$
\\ 
\textcolor{red}{Exact value}  & \textcolor{red}{0.0360} & - & -& \textcolor{red}{0.360} & -&-\\ \bottomrule
\end{tabular}}
\end{table}
\subsection{ Boundary Heat Flux Identification}

In this section, we apply our framework to study an inverse heat conduction problem (IHCP) where boundary conditions are partially accessible. Typically, these problems arise in a plethora of industrial and engineering applications where measurements can only be made in easily-accessible locations or the quantity of interest can be measured indirectly. Unfortunately, inverse problems are ill-posed and ill-conditioned because unknown solutions and parameter values usually have to be determined from indirect observable data that contains measurement error \cite{beck1985inverse,hon2004fundamental,shen1999numerical,wang2004bayesian}. Here, we aim to identify spatio-temporal boundary heat flux given partial spatio-temporal temperature observations inside the domain as in the work of \citet{wang2004bayesian}.

\begin{subequations}
\begin{align}
    &\frac{\partial T}{\partial t} = \frac{\partial^2 T}{\partial x^2 } + \frac{\partial^2 T}{\partial y^2 }, ~ 0< x,y< 1, t \in [0,1],\\
    &T(x,y,0) = -2 \sin(\pi x) \sin(\pi y) , ~ 0 \le x,y \le 1, \\
    &T|_{x=1} = T|_{y=1} = 0, ~ 0 < t< 1 \\
    &\frac{\partial T}{\partial x}\Bigg|_{x=0} = q_x~\text{(unknown)},~ 0< t < 1, \\
    &\frac{\partial T}{\partial y}\Bigg|_{y=0} = q_y,~\text{(unknown)} ~ 0< t < 1,
\end{align}
\end{subequations}
  
where $q_x $ and $q_y$ are the unknown heat fluxes to be discovered. As considered in \cite{wang2004bayesian}, an analytical solution to this problem can be obtained as follows 
\begin{equation}
    T(x,y,t) = -2 \pi \sin(\pi x) \sin (\pi y) e^{-2 \pi^2 t},
\end{equation}
with the exact heat fluxes as follows 
\begin{subequations}
\begin{align}
    q_x &=  -2\pi \sin(\pi y) e{-2\pi^2 t}\\
    q_y &= -2\pi \sin(\pi x) e{-2\pi^2 t}.
\end{align}
\end{subequations}
 
\begin{figure}[!ht]
    \centering
    \subfloat[]{\includegraphics[width=0.25\textwidth]{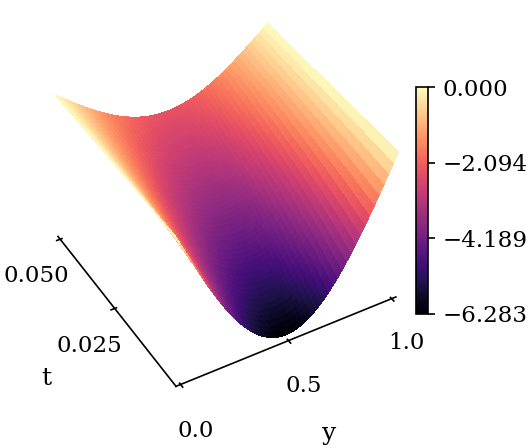}}\hspace{3em}
    \subfloat[]{\includegraphics[width=0.25\textwidth]{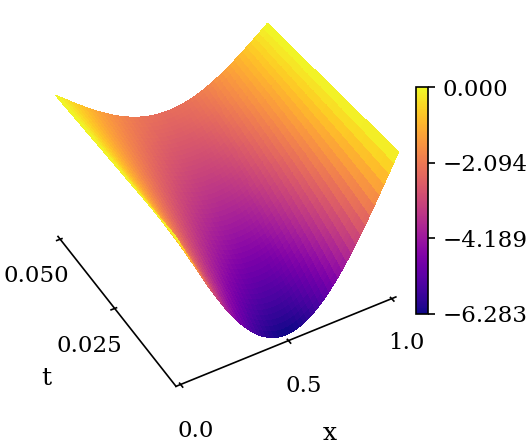}}\hspace{3em}
    \subfloat[]{\includegraphics[width=0.23\textwidth]{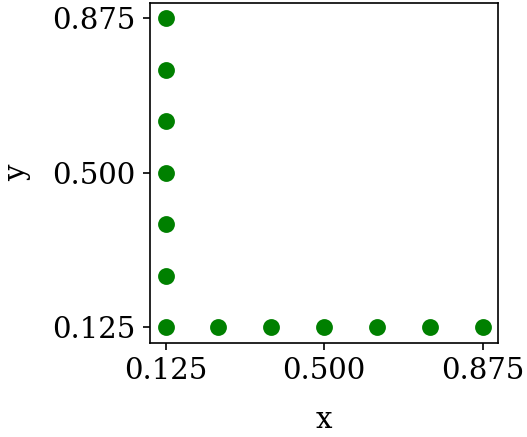}}
    \caption{ (a) exact $q_x$, (b) exact $q_y$, (c) location of thermocouples}
    \label{fig:ExactHeatFluxes}
\end{figure}

An exact representation of $q_x$ and $q_y$ are presented in Fig.~\ref{fig:ExactHeatFluxes}(a) and (c). The inverse problem is to discover $q_x$ and $q_y$ given partial observation from a set of thirteen thermocouples with $0.125$ space interval and $0.125$ distance to the boundary as shown in Fig.~\ref{fig:ExactHeatFluxes}(c). 

\begin{figure}[!ht]
    \centering
    % predictions qx
    \subfloat[]{\includegraphics[width=0.25\textwidth]{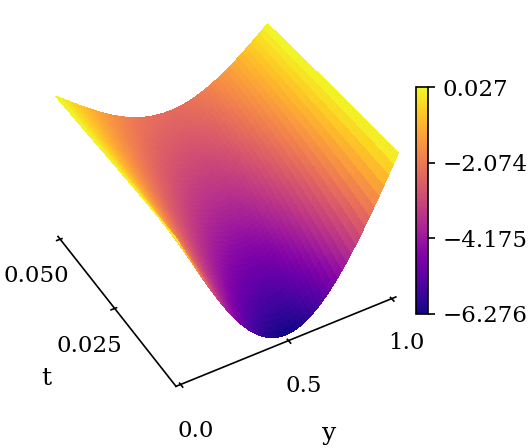}}\hspace{3em}
    \subfloat[]{\includegraphics[width=0.25\textwidth]{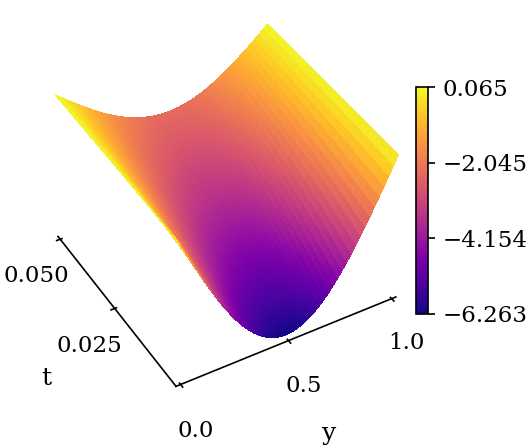}}\hspace{3em}
    \subfloat[]{\includegraphics[width=0.25\textwidth]{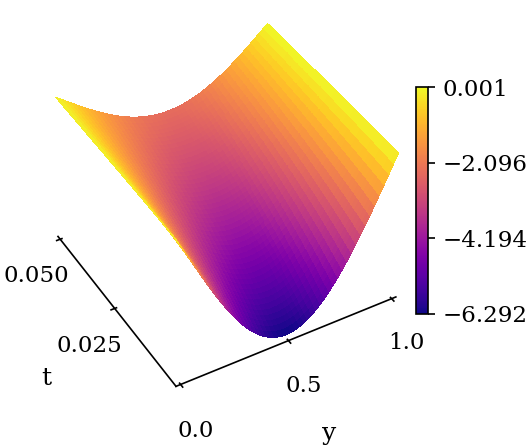}}\\
    % predictions qy
     \subfloat[]{\includegraphics[width=0.25\textwidth]{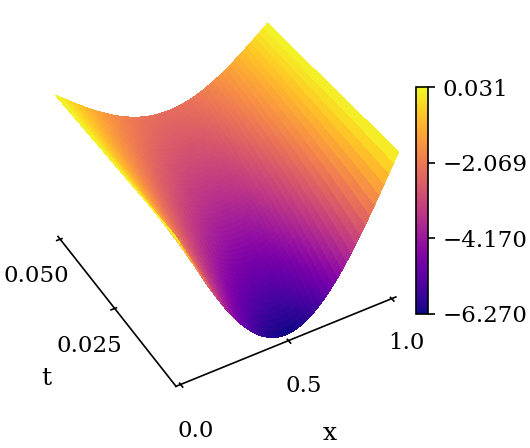}}\hspace{3em}
    \subfloat[]{\includegraphics[width=0.25\textwidth]{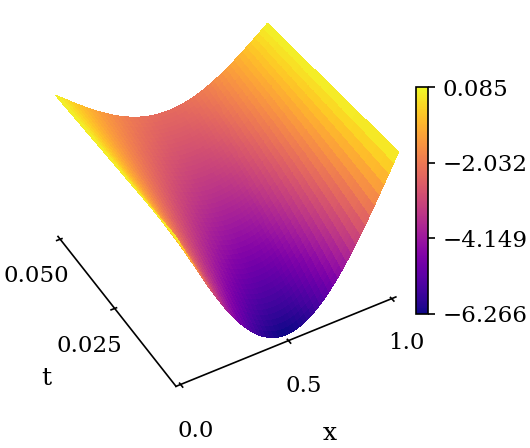}}\hspace{3em}
    \subfloat[]{\includegraphics[width=0.25\textwidth]{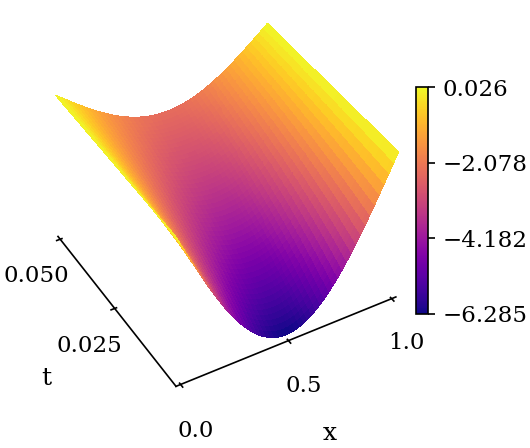}}
    \caption{Heat flux reconstruction: Top row: $q_x$ (a) predicted flux distribution for case I, (b) predicted flux distribution for case II, (c) predicted flux distribution for case III. Bottom row: $q_y$, (d) predicted flux distribution for case I, (e) predicted flux distribution for case II, (f) predicted flux distribution for case III. }
    \label{fig:flux_reconstruction}
\end{figure}

The sampling time interval is taken as $dt = 0.002$. The heat flux history was reconstructed for the time range $t \in [0:0.05]$,  N = 25, hence, there are 325 observations. \citet{wang2004bayesian} represented the unknown flux quantities by parametric linear functions and proposed a Bayesian approach by employing a specialized model of Markov random field (MRF) as prior distribution. Three different cases were considered. Uncertainty in temperature measurements was modeled as stationary zero-mean white noise with standard deviations of $\sigma = 0.005$ , $\sigma = 0.01$ and $\sigma = 0.02$. 
We employ a 3 hidden-layer fully-connected neural network with 30 neurons per layer to learn the temperature field for the entire domain. Our optimizer is LBFGS with its default parameters and \emph{strong-wolfe} line search function built-in PyTorch framework. We set the limiting penalty parameter $\mu_{\max} = 10^{4}$ similar to previous problems and we train our network for 10000 epochs. We use Sobol sequences to sample $512$ residual points in the domain, $512$ points for the Dirichlet boundary conditions, and $512$ points for the initial condition only once before training our network. The predictions of our neural network model are shown in Fig.~\ref{fig:flux_reconstruction}. We observe that our network has successfully inferred heat fluxes for all three cases. A summary of the error percentage from our method along with the best results from \cite{wang2004bayesian} are provided in Table~\ref{tb:heat_flux}. We observe that our approach has improved the reported results of \cite{wang2004bayesian} by a factor of $10$ in all three cases. 

\begin{table}[ht!]
\centering
\caption{$q_x$ reconstruction error by different methods with noisy measurement data}
\label{tb:heat_flux}
\vspace{2pt}
\resizebox{0.5\textwidth}{!}{%
\begin{tabular}{@{}lcccc@{}}
\toprule
\multicolumn{1}{c}{Models} &
  \multicolumn{1}{c}{$\sigma = 0.005$} &
  \multicolumn{1}{c}{$\sigma = 0.01$} &
  \multicolumn{1}{c}{$\sigma = 0.02$} 
  \\ \midrule
Ref. \cite{wang2004bayesian}  &  4.62\%         & 5.45\%        & 5.75\%  \\
PECANN with MF data  & $\boldsymbol{0.53\%}$         & $\boldsymbol{0.61\% }$       & $\boldsymbol{0.89\%  }$   
\\  \bottomrule
\end{tabular}}
\end{table}
\subsection{Patient-specific Tumor Growth Modeling}

\begin{figure}[!ht]
\centering
    \subfloat{\includegraphics[width=0.25\textwidth]{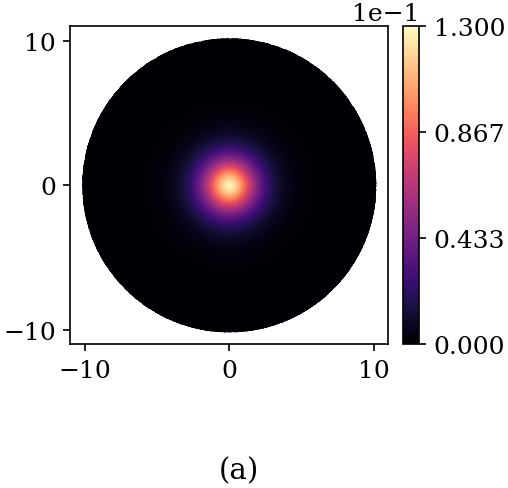}}\hspace{4em}
    \subfloat{\includegraphics[width=0.25\textwidth]{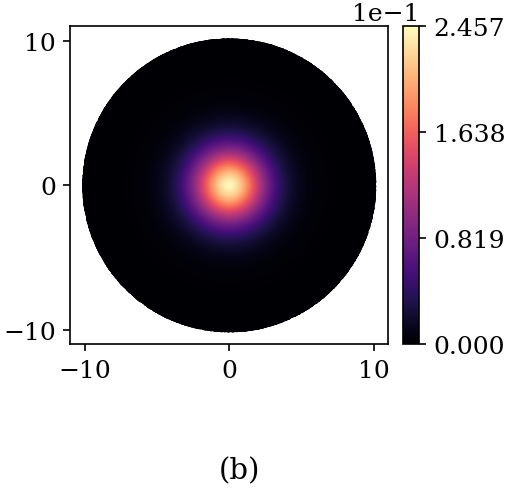}}\hspace{4em}
    \subfloat{\includegraphics[width=0.25\textwidth]{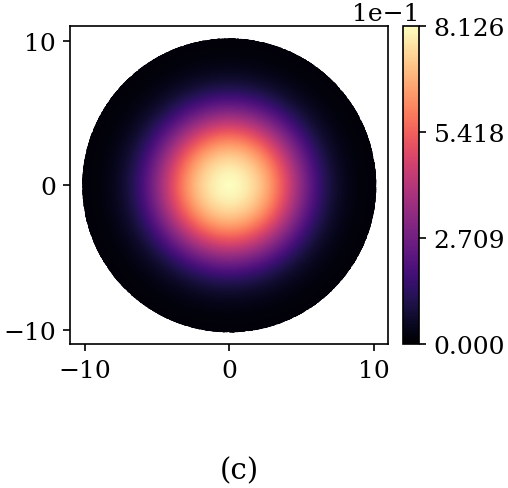}}\\
    \subfloat{\includegraphics[width=0.25\textwidth]{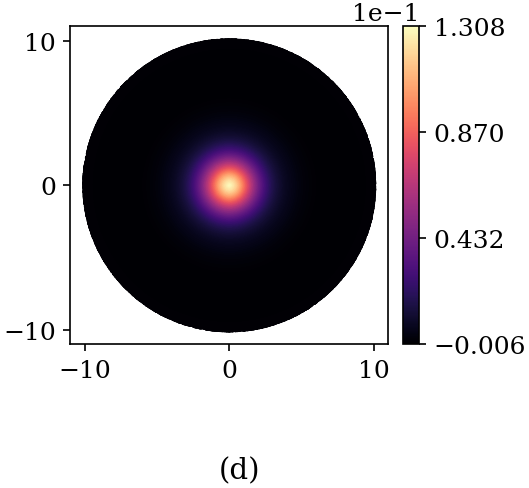}}\hspace{4em}
    \subfloat{\includegraphics[width=0.25\textwidth]{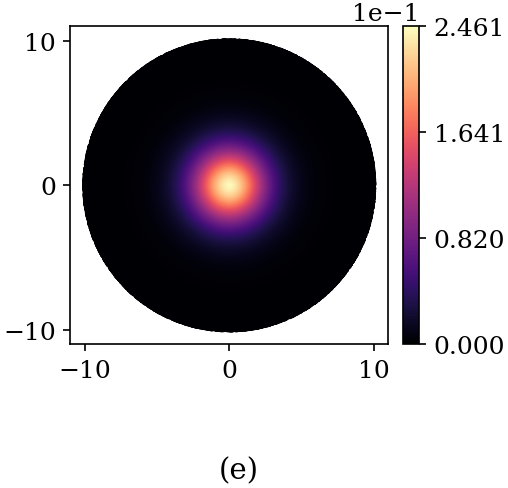}}\hspace{4em}
    \subfloat{\includegraphics[width=0.25\textwidth]{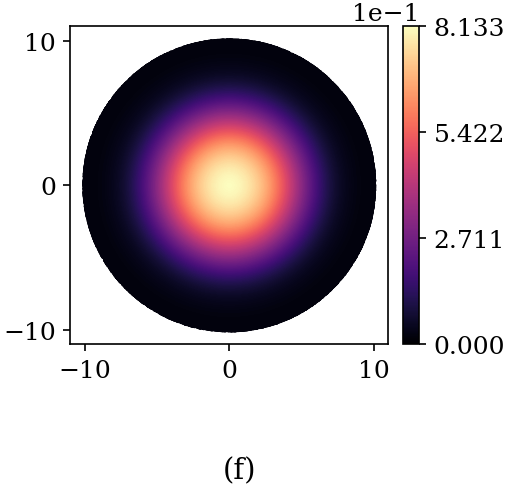}}
    \caption{Natural tumor cell density distribution at different time states. Top row: (a) synthetic brain scan at year one used for training, (b) synthetic brain scan at year two used for training,  (c) synthetic brain scan at year five used for testing . Bottom row: (d) reconstructed brain scan data at year one, (e) reconstructed brain scan data at year two, (f) predicted brain scan data at year five. }
    \label{fig:tumor_results}
\end{figure}

In this section, we aim to develop a patient-specific tumor model using noisy magnetic resonance images (MRI). Treatment for tumors involves surgery, radiation, and chemotherapy. Nevertheless, cancer cells may remain after surgery, resulting in recurrence of the tumor and eventual death \cite{swanson2003virtual,giese2003cost}. Therefore, models based on patient-specific information are needed to identify tumor cells that may lie beyond the threshold visible to magnetic resonance imaging. Assuming isotropic brain structure and radial symmetry, we can describe tumor cell density evolution using the following non-linear reaction-diffusion type partial differential equation \cite{jaroudi2018inverse,jbabdi2005simulation,painter2013mathematical,ozuugurlu2015note,rockne2009mathematical}.
\begin{align}
    \frac{\partial u(r,t)}{\partial t }  &= D \frac{\partial^2 u(r,t)}{\partial r^2} + \rho u(r,t)(1-u(r,t)), ~ \text{in}~ \Omega \times [0,5]
    \label{eq:reaction_diffusion}\\
    \frac{\partial u(r,t)}{\partial r} &= 0, ~\text{on}~ \partial \Omega
    \label{eq:reaction_diffusion_bc}\\
    u(r,0) &= \varphi(r), ~\text{in}~ \Omega
    \label{eq:reaction_diffusion_ic}
\end{align}
where $\Omega = \{r ~|~ 0 \le r \le 10\}$ is the domain with its boundary $\partial \Omega$, $u(r,t)$ is the unknown tumor cell density at time $t$ [year] and distance $r$ [mm]. $D$ is the unknown diffusion coefficient of tumor cells in the brain tissue and $\rho$ is the unknown proliferation coefficient. $\varphi$ is a point source initial condition. It is assumed that at the time of death $t=5$, the visually detectable area of tumor volume is equal to a circle of 10 mm in radius. As a proof of concept, we generate synthetic MRI data by solving Eq.~\eqref{eq:reaction_diffusion} in forward mode using finite difference scheme with $\Delta r=0.0196, \Delta t = 10^{-5}$ assuming $D = 0.50 $, $\rho = 1.00$ with the following initial condition
\begin{align}
    \varphi(r)  = \frac{1}{10}e^{-r}, ~\text{in}~\Omega.
\end{align}
Our synthetic data includes two solutions at $t=1$ and $t=2$ that simulate patient tumor cell density distribution obtained from MRI of brain scans at the corresponding time states. We further corrupt these data using uncorrelated Gaussian noise with $\sigma = 0.01$. The corresponding noise percent of the data is presented in Table~\ref{tb:MRI_noise}. 
% noise percent in training MRI data 
\begin{table}[!ht]
\caption{Error percentage of corrupted MRI data with uncorrelated Gaussian noise with standard deviation $\sigma = 0.01$}
\label{tb:MRI_noise}
\begin{center}
\begin{tabular}{@{}lcc@{}}
\toprule
& $u(r,t=1)$ & $u(r,t=2)$ \\ \midrule
Noise $\%$ & $10.34 \%$         & $5.01 \%$ \\ \bottomrule
\end{tabular}
\end{center}
\end{table}
From Table~\ref{tb:MRI_noise} we observe that our data contain different levels of noise which indicates different levels of fidelity. Finally, we use our corrupted tumor density distribution at $t=1$ and $t=2$ as low fidelity data along with Eq.~\eqref{eq:reaction_diffusion_bc} as our boundary constraint (high fidelity data) to infer unknown parameters of Eq.~\eqref{eq:reaction_diffusion}. For this problem, we generate $N_{\Omega} = 512$ residual points to approximate the loss on Eq.~\eqref{eq:reaction_diffusion} and $N_{\partial \Omega} = 512$ to constrain the boundaries only once before training. We also generate $512$ points with their labels from our corrupted synthetic brain scan data. We use a feed-forward neural network with two hidden layers each with 10 neurons per layer. Our optimizer is LBFGS with its default parameters and \emph{strong wolfe} line search function built in PyTorch framework \cite{paszke2019pytorch}. Our network is trained for 200 epochs with Xavier initialization scheme \cite{glorot2010understanding}. We initialize $D \in [0.3285, 0.973]$ and $\rho \in [0.73, 2.92]$ randomly as suggested in \cite{larsson2019solving}. 
From Fig.~\ref{fig:tumor_results} we observe that our model not only reconstructs the original data from corrupted noisy data, but also generalizes well to predict the unseen MRI data at the terminal year $t=5$. 
A summary of our inferred parameters are presented on table \ref{tab:Inferred_Patient_Data} over 10 independent trials with random Xavier initialization scheme \cite{glorot2010understanding}. 
% inferred parameters table
\begin{table}[!hb]
\caption{Tumor growth modeling. Summary of inferred parameters using multi-fidelity data in the learning process averaged over ten independent trials with random initialization}
\label{tab:Inferred_Patient_Data}
\begin{center}
\resizebox{0.5\textwidth}{!}{
\begin{tabular}{@{}lcc@{}}
\toprule
& D & $\rho$ \\ \midrule
Exact &$0.50$ & $1.00$\\
PECANN with MF data & $0.49 \pm 4.90 \times 10^{-3}$ &$1.00 \pm 9.93\times 10^{-4}$ \\
\bottomrule
\end{tabular}
}
\end{center}
\end{table}

\section{Conclusion}
\label{sec:Conclusion}
We have shown that the unconstrained optimization problem formulation pursued in physics-informed neural networks (PINN) is a major source of poor performance when the PINN approach is applied to learn the solution of more challenging multi-dimensional PDEs. We addressed this issue by introducing physics- and equality-constrained artificial neural networks (PECANN), in which we pursue a constrained-optimization technique to formulate the objective function in the first place. Specifically, we adopt the augmented Lagrangian method (ALM) to constrain the PDE solution with its boundary and initial conditions, and with any high-fidelity data that may be available. The objective function formulation in the PECANN framework is sufficiently general to admit low-fidelity data to regularize the hypothesis space in inverse problems as well. We applied our PECANN framework for the solution of both forward problems and inverse problems with multi-fidelity data fusion. For all the problems considered, the PECANN framework produced results that are in excellent agreement with exact solutions, while the PINN approach failed to produce acceptable predictions.

It is a common practice to use conventional feed-forward neural networks in the PINN approach. However, these type of neural networks are known to suffer from the so-called vanishing gradient problem, which stalls the learning process. Residual layers (a.k.a. ResNets) that were originally proposed by \citet{he2016deep} tackle the vanishing gradient problem with identity skip connections. In our work, we have modified the original residual layers by restricting them to a single weight layer with a $tanh$ activation function and identity skip connections. We find our leaner version of the residual layers to be very effective in improving the accuracy of the PDE predictions for both the original PINN model and our PECANN model.

Our findings suggest that not only the choice of the neural network architecture, but also the optimization problem formulation is crucial in accurately learning PDEs using artificial neural networks. We conjecture that future progress in physics-constrained (informed) learning of PDEs would come from exploring new approaches in the field of non-convex constrained optimization field. Future endeavors could shed light on challenging questions such as: how does the loss landscape of neural networks change with respect to the optimization problem formulation? What is the optimal neural network architecture for PDE learning? And, is there a physics-based approach in searching for optimal architectures? 
\section*{Acknowledgments}
This material is based upon work supported by the National Science Foundation under Grant No. 1953204 and in part in part by the University of Pittsburgh Center for Research Computing through the resources provided.

\appendix
\section{Additional Examples}\label{sec:appendix}
\subsection{One-dimensional Poisson's Equation}
The aims of this pedagogical example are twofold: First, we thoroughly demonstrate the implementation intricacies of our proposed method and highlight its advantages over the PINN approach. Second, we demonstrate the significant improvement achieved in the model prediction using our modified residual architecture relative to the original residual networks \cite{he2016deep}.  

Let us consider the following one-dimensional Poisson's equation
\begin{align}
    &\frac{d^2 u}{dx^2} = -(15\pi )^2 \cos(15\pi x) , &&x \in \Omega ,\\
    &u(x) = \cos(15 \pi x),   &&x \in \partial \Omega,
    \label{eq:Poisson_1D}
\end{align}
where $\Omega = \{x \hspace{0.15em} | \hspace{0.15em} 0 \leq x \leq 1\}$ and $\partial \Omega$ is its boundary. The exact solution to the above problem is a sinusoidal nonlinear function $u(x) = \cos(15 \pi x)$. Considering a neural network solution for the above equation as $\hat{u}(x;\theta)$ parameterized with $\theta$, we write the residual form of this one-dimensional Poisson's equation as follows:
\begin{align}
    \mathcal{F}&:= \frac{d^2 u_{\theta} }{dx^2}+ (15\pi )^2 \cos(15\pi x) &&x \in \Omega,    \label{eq:residualFormOfOnedimensionalPoisson} \\
    \mathcal{B}&:= u_{\theta}(x)  - \cos(15 \pi x), &&x \in \partial \Omega.
    \label{eq:residualFormOfOnedimensionalPoissonBC}
\end{align}
Next, we use the above residual form of this differential equation to construct an objective function as proposed earlier in Eq.\eqref{eq:ProposedObjectiveFunction}
\begin{equation}
    \mathcal{L}(\theta)  =\sum_{i=1}^{N_{\Omega}}|\mathcal{F}(x^{(i)})|^2
    + \sum_{i=1}^{2}\lambda^{(i)}\phi(\mathcal{B}(x^{(i)}))
    + \frac{\mu}{2}\sum_{i=1}^{2}|\phi(\mathcal{B}(x^{(i)}))|^2
    \label{eq:Poisson_PECANN}
\end{equation}
where $\mu$ is the penalty parameter and  $N_{\Omega}$ is the number residual points sample from $\Omega$ at every epoch. $\lambda \in \mathbb{R}^2$ is a vector of Lagrange multipliers for the boundary constraints and $\phi$ is the quadratic distance function. In contrast to our constrained optimization with the ALM, the composite objective function adopted in PINNs (i.e. Eq.~\ref{eq:PINN_loss}) yields the following loss function for the current example
\begin{equation}
    L(\theta) = \frac{1}{N_\Omega}
    \sum_{i=1}^{N_{\Omega}}|\mathcal{F}(x^{(i)})|^2
    + 
    \frac{1}{2}
    \sum_{i=1}^{2}|\mathcal{B}(x^{(i)})|^2
    \label{eq:Poisson_PINN}
\end{equation}
Having constructed the objective functions using the constrained-optimization method in the present work and the composite approach adopted in PINNs, we design three networks in such a way that they have the same number of neurons and hidden layers to allow a fair comparison. 
We use six weight layers with 50 neurons per layer in all three neural network models. More specifically, we have six weight layers in our conventional feed-forward neural network model. Similarly, our second neural network model with the original residual layers has one weight layer in the front with two residual layers and an output weight layer, which makes a total of six weight layers. For our last neural network model with our proposed residual layers, we have a weight layer succeeded by four modified residual layers and an output weight layer that amounts to six weight layers as well. Therefore, all three models have the same number of neurons and the same number of weight layers and are end-to-end trainable. For this problem, the parameters of the network are initialized randomly with the Xavier initialization technique \cite{glorot2010understanding}. We use Adam \cite{kingma2014adam} with an initial learning rate of $10^{-2}$. We reduce our learning rate by a factor of $0.95$ after 100 epochs with no improvement in the objective function. We use the same hyperparameters and train all the models under the same training settings with both objectives as in Eq.~\eqref{eq:Poisson_PECANN} and Eq.~\eqref{eq:Poisson_PINN}. We set the limiting penalty parameter $\mu^{\max} = 10^2$. As for the collocation points, we randomly generate $N_{\Omega} =654$ residual points from across our domain with uniform probability along with two boundary conditions at each optimization step.
% Figure (1) : Performance comparison
\begin{figure}[t]
    \centering
    \subfloat{\includegraphics[width=\textwidth]{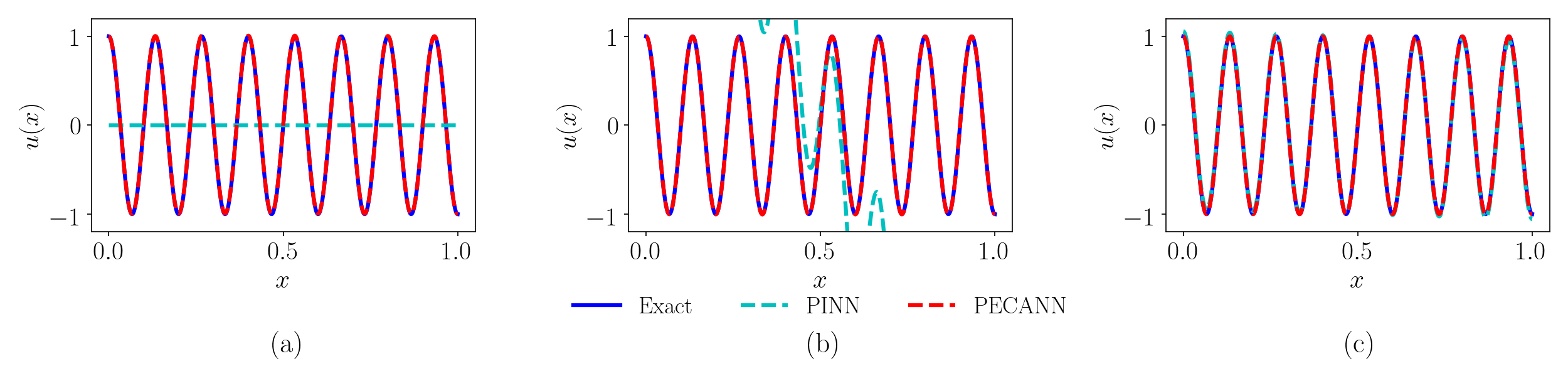}}
    \caption{Performance comparison of PINN vs PECAN for different neural network architectures: (a) conventional neural network, (b) original residual neural network, (c) modified residual neural network. Note that PECANN approach converged with all network architectures while PINN only converged with our proposed modified residual neural network but with poor norms of errors }
    \label{fig:SolutionComparisonPoisson_1D}
\end{figure}
The results from all three neural network architectures trained with the PINN and PECANN approaches are juxtaposed in Fig.~\ref{fig:SolutionComparisonPoisson_1D}. We observe from these results that the PINN model with a composite objective function is visibly sensitive to the neural network choice and benefits the most from the adoption of modified residual layers, whereas the PECANN model with equality-constrained optimization is qualitatively less sensitive to the choice of the neural network architecture and performs very well for all three networks.
\begin{table}[!ht]
\centering
\caption{One dimensional Poisson's equation: summary of relative $L_2$ norms and $L_\infty$ norms for different neural network models trained with different approaches}
\label{tb:Poisson_1D}
\resizebox{0.70\textwidth}{!}{%
\begin{tabular}{lcccc}
\hline
 & \multicolumn{2}{c}{PINN} & \multicolumn{2}{c}{PECANN} \\ \hline
 & \multicolumn{1}{c|}{Relative $L_2$} & \multicolumn{1}{c|}{$L_{\infty}$} & \multicolumn{1}{c|}{$L_2$} & $L_{\infty}$ \\ \hline
\multicolumn{1}{l|}{Conventional NN} & $1.00$ & $1.00$ & $\boldsymbol{1.73 \times 10^{-3}}$ & $\boldsymbol{1.80 \times 10^{-3}}$ \\
\multicolumn{1}{l|}{Original Residual NN} & $4.75$ & $5.84$ & $\boldsymbol{3.41 \times 10^{-3}}$ & $\boldsymbol{3.43 \times 10^{-3}}$ \\
\multicolumn{1}{l|}{Proposed Residual NN} & $4.63 \times 10^{-2}$ & $5.74\times 10^{-2}$ & $\boldsymbol{1.35 \times 10^{-4}}$ & $\boldsymbol{1.66 \times 10^{-4}}$
\end{tabular}
}
\end{table}
From Table~\ref{tb:Poisson_1D} 
we observe that for conventional NN and for the original residual neural network the relative $L_2$ error from our PECANN model is three orders of magnitude lower than the one obtained from our PINN model. However, with our lean residual network, it decreases to two orders of magnitude which demonstrates the impact of our neural network architecture. 
\subsection{Three-dimensional Poisson's Equation}
\begin{figure}[!ht]
    \centering
    \subfloat{\includegraphics[width=\textwidth]{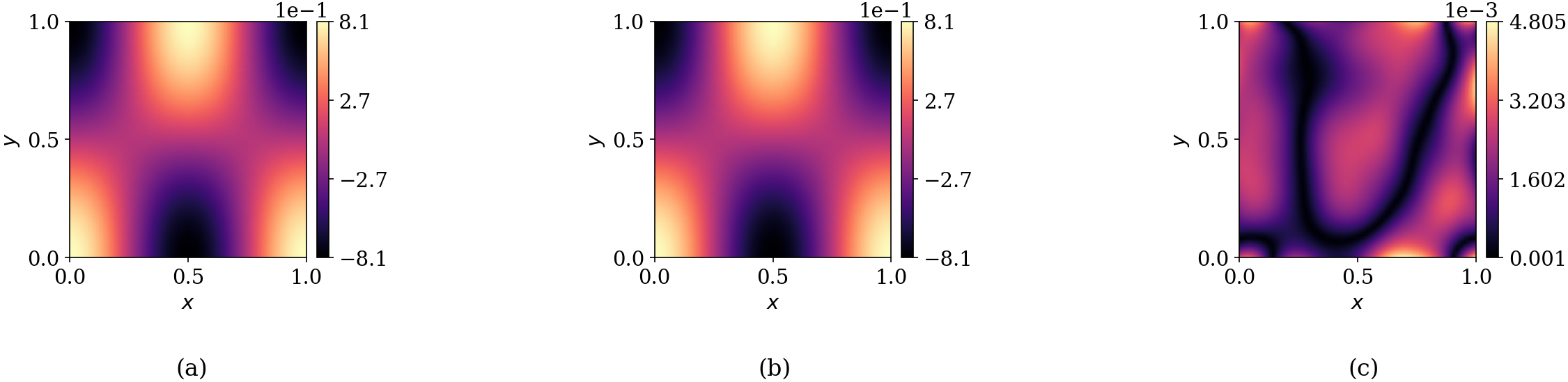}}\\
    \subfloat{\includegraphics[width=0.98\textwidth]{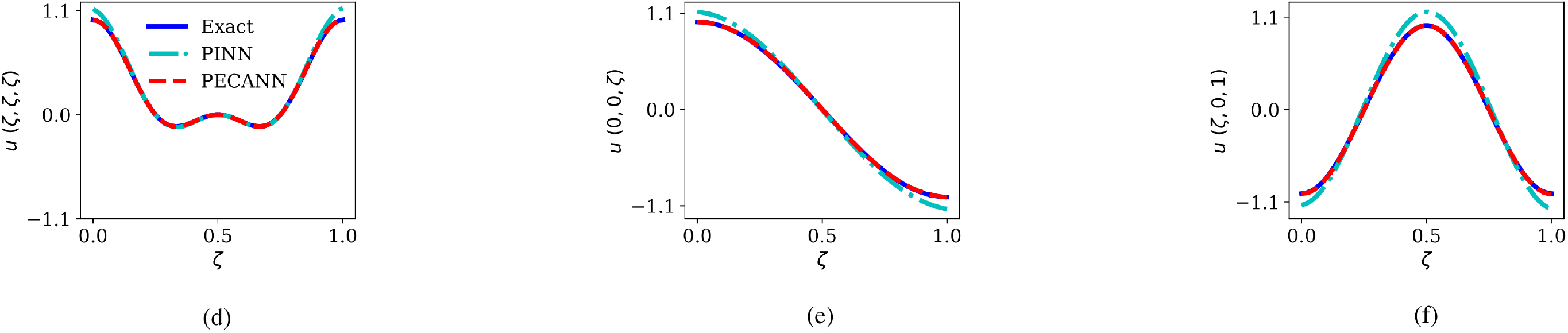}}
    \caption{Three dimensional Poisson's equation. Top row: cross section view of the solution at $z = 0.2$,  (a) exact solution ,(b) predicted solution from PECANN model, (c) absolute point-wise error distribution. Bottom row: plots over line obtained from different methods, (d) straight line connecting point $(0,0,0)$ to point $(1,1,1)$, (e) straight line connecting $(0,0,0)$ and $(0,0,1)$ points, (f) straight line connecting point $(0,0,1)$ to point at $(1,0,1)$}
    \label{fig:Poisson_3D}
\end{figure}
We consider the following non-homogeneous three dimensional Poisson's equation in a cubic domain 
\begin{equation}
    \nabla^2 u(x,y,z)=f(x,y,z), \qquad (x,y,z)\in \Omega,
    \label{eq:threeDimPoisson}
\end{equation}
subject to the following boundary conditions 
\begin{equation}
    u(x,y,z)=g(x,y,z), \qquad (x,y,z) \in \partial \Omega,
    \label{eq:threeDimPoissonBC}
\end{equation}
where $\Omega = \{0 \le x,y,z \le 1 \}$ with its boundary $\partial \Omega$, $f$ and $g$ are known source functions in $\Omega$ and on $\partial \Omega$ . We manufacture a sinusoidal solution of the following form
\begin{equation}
    u(x,y,z) = \cos ( 2\pi x) \cos(\pi y) \cos(\pi z), \forall (x,y,z) \in \Omega,
    \label{eq:Exact_Poisson_3D}
\end{equation}
We will use the exact solution \cref{eq:Exact_Poisson_3D} to evaluate the source functions $f$ and $g$ and solve Eq.\eqref{eq:threeDimPoisson}. For this purpose, we use our lean residual neural network with three hidden layers each with 50 neurons. Our optimizer is Adam \cite{kingma2014adam} with its default parameters and $10^{-2}$ initial learning rate. We also reduce our learning rate by a factor of $0.95$ if the objective does not improve after 100 optimization steps. Our network is trained for $15000$ epochs with randomly initialized weights according to Xavier scheme \cite{glorot2010understanding}. We generate $N_{\partial \Omega} = 6\times 256$ number of points on the boundaries $\partial \Omega$ only once before training and $N_{\Omega} = 256$ residual points in the domain $\Omega$ at every optimization step. 
We present a section view of the predicted solution by our PECANN model in Fig.~\ref{fig:Poisson_3D}. 
\begin{table}[ht!]
\centering
\caption{Three dimensional Poisson's equation: summary of relative $L_2$ norms and $L_\infty$ norms for different neural network models averaged over 10 independent trials with random initialization with Xavier scheme}
\label{tb:Poisson3D}
\vspace{2pt}
\resizebox{0.80\textwidth}{!}{%
\begin{tabular}{@{}lcccccc@{}}
\toprule
\multicolumn{1}{c}{Models} &
  \multicolumn{1}{c}{Relative $L_{2}$} &
  \multicolumn{1}{c}{$L_\infty$} &
    \multicolumn{1}{c}{$N_{\Omega}$} &
  \multicolumn{1}{c}{$N_{\partial \Omega}$} &
  \\ \midrule
PINN  & $1.09 \times 10^{-1} \pm 1.54 \times 10^{-2}$ & $2.31 \times 10^{-1} \pm 4.56 \times 10^{-2}$& $256 \times 15000 $&$6 \times 256$\\
PECANN & $\boldsymbol{2.39 \times 10^{-3} \pm 2.93 \times 10^{-4} }$& $\boldsymbol{6.21 \times 10^{-3} \pm 1.55 \times 10^{-3}}$&  $256 \times 15000 $&$6 \times 256$
\end{tabular}}
\end{table}
Since our physics-informed neural network failed to converge as can be seen from the error norms in Table \ref{tb:Poisson3D}, we did not include a section view of its predicted solution. However, we provide plots over straight lines drawn between two points within the domain. From Fig.~\ref{fig:Poisson_3D}(c)-(e) we observe that our PINN model either underpredicted or overpredicted the regions with high gradients and regions close to the boundaries. However, our PECANN model successfully learned the underlying solution. 
From Table~ \label{tb:Poisson_3D} 
we observe that for the relative $L_2$ error from our PECANN model is two orders of magnitude lower than the one obtained from our PINN model which highlights the effectiveness of our method over conventional physics-informed neural networks.

\bibliographystyle{plainnat}
\bibliography{citations}
\end{document}